\documentclass[12pt]{article}
\usepackage{amsmath, amssymb, mathrsfs, mathtools, tensor, braket}
\usepackage{xcolor, graphicx, subfigure}
\usepackage{cite, varioref}
\numberwithin{equation}{section}

\usepackage{enumitem, float}
\usepackage{caption}
\evensidemargin \oddsidemargin

\begin{document}

\begin{titlepage}

\begin{center}
\vspace{5mm}
   
% Title
{\Large \bfseries
Charged Vortex in Superconductor
}\\[17mm]
% Authors
Yoonbai Kim,
~~SeungJun Jeon,
~~Hanwool Song
\\[3mm]  
% Departments & E-mail
{\itshape
Department of Physics,
Sungkyunkwan University,
Suwon 16419,
Korea
\\[-1mm]
yoonbai@skku.edu,~
sjjeon@skku.edu,~
hanwoolsong0@gmail.com
}
\end{center}
\vspace{15mm}

\begin{abstract}
We find the charged spinless vortices in the effective field theory of a Schr\"{o}dinger type complex scalar field of Cooper pair, a U(1) gauge field of electromagnetism, and a gapless neutral scalar field of acoustic phonon. We show that regular static vortex solutions are obtained only for the nonzero critical cubic Yukawa type coupling between neutral and complex scalar fields. Since the Coulombic electric field is exactly cancelled by the phonon, the obtained charged vortices have finite energy. 
When the quartic self-interaction coupling of complex scalar field has the critical value, the BPS (Bogomolny-Prasad-Sommerfield) bound is saturated for multiple charged vortices of arbitrary separations and hence the borderline of type I and I$\!$I superconductors is achieved in nonperturbative 
regime.
\end{abstract}

\end{titlepage}

\section{Introduction}

Theory of conventional superconductivity \cite{Arovas:2019,tinkham2004introduction, Bennemann:2008} is the Bardeen-Cooper-Schrieffer (BCS) theory in electron level \cite{Bardeen:1957mv} while, in the level of direct treatment of the Cooper pair, more effective description is made by the Ginzburg-Landau theory \cite{Ginzburg:1950sr}. In the framework of the Ginzburg-Landau theory, a prominent result is the existence of vortices and the classification of type I and I$\!$I superconductors \cite{Abrikosov:1956sx}.

A characteristic property of the Abrikosov-Nielsen-Olesen vortices is electrical neutrality. Though it seems less clear in the Ginzburg-Landau free energy for only time-independent configurations, it becomes obvious in the Abelian Higgs model with minimal electromagnetic coupling as far as static configurations are concerned \cite{Nielsen:1973cs}. Therefore, a long time attempt has been tried in order to find charged vortices in a natural way in the context of U(1) gauge theories \cite{Careri:1965}.
Charged vortices are obtained mostly by charge trapping and extensive efforts have been made in high temperature superconductivity  \cite{Hagen:1991,Khomskii:1995,Blatter:1996,Nagaoka:1998,Kumagai:2001, Jan:2001,Mounce:2011,Sahu:2022}.
In this work, we focus on the field theory of superconductivity of $s$-waves including a nonrelativistic Schr\"{o}dinger type complex scalar field and .

Zero or nonzero U(1) charge is governed by the Gauss' law. Hence an easy way is to change its gauge dynamics side of $\nabla\cdot\boldsymbol{E}$. For example, the Chern-Simons solitons carry U(1) charge
\cite{Hong:1990yh,Jackiw:1990aw} but their low energy dynamics is governed not by the Maxwell theory of electromagnetism but by the $(1+2)$-dimensional Chern-Simons term of linear derivative. Therefore, we will avoid change of gauge dynamics but strictly keep electromagnetism by solely by the Maxwell term. The other way is to introduce a different form of charge density $\rho$. As the complex scalar field of slowly moving Cooper pairs is concerned in superconducting samples, the first candidate must be to take into account the Schr\"{o}dinger type nonrelativistic matter.\footnote{There is another way to consider nonrelativistic complex scalar field with the help of characteristic propagation speed of sound waves
\cite{Kim:2024gfn,Jeon:2025snd}. However, the Gauss' law is unaltered and the obtained static vortices of finite energy are left to be electrically neutral.} Once the field theory of electromagnetism and nonrelativistic Schr\"{o}dinger type complex scalar field is considered with the minimal electromagnetic coupling and the coupling of quartic scalar self-interaction accompanied with spontaneous symmetry breaking, the vortices have electric charge but not finite energy. Even nonexistence of nonsingular static vortex solution is expected and hence we will address this question in section 4.

In the previous works \cite{FTSC, NRAH}, we proposed an effective field theory added by a gapless neutral scalar field of acoustic phonon with a cubic Yukawa type coupling to the complex scalar field of Cooper pair.
Here we show that the field theory of our consideration supports the finite energy solutions of charged spinless vortex characterized by quantized magnetic flux only when the cubic Yukawa type coupling has a critical value for superconducting phase.
In addition, when the quartic self-interaction coupling takes a critical value, we find the Bogomolny-Prasad-Sommerfield (BPS) bound \cite{Bogomolny:1975de} saturated by the charged but noninteracting vortices. 
Since the obtained critical coupling coincides with that of the neutral Abrikosov-Nielsen-Olesen vortices \cite{Abrikosov:1956sx,Bogomolny:1975de}, the charged vortices attract each other in weak coupling regime of type I superconductors and repel in strong coupling regime of type I$\!$I superconductors. 
Existence of such BPS limit means nonperturbative confirmation of the interaction balances in the static limit, established previously in perturbative regime \cite{NRAH}.

The rest of this paper is organized as follows.
In section 2, we consider the effective field theory of interest and recapitulate briefly perturbative aspects.
In section 3, we derive the BPS bound for charged noninteracting vortices with quantized magnetic flux and electric charge but without spin.
In section 4, we show that the static charged vortices of finite energy are obtained only for the critical cubic Yukawa type coupling and study detailed properties of charged vortices.
We conclude in section 5 with discussions.

\section{Effective Field Theory of Superconductivity and Perturbative Aspects}

In this section, we introduce the effective field theory for superconductivity of $s$-wave and review its superconducting vacuum and perturbative aspects~\cite{FTSC, NRAH}.
Field theoretic description of conventional superconductors is proposed by writing the action possessing a U(1) gauge field $A^{\mu}=(\Phi/c,A^{i})$ ($\mu= t, 1, 2, 3$ and $i= 1, 2, 3$)  of electromagnetism, a complex scalar field $\Psi$ of mass $m=2m_{{\rm e}}$ and charge $q=-2e$ at tree level for Cooper pair, and a neutral scalar field $N$ of an acoustic phonon of propagation speed $v_{N}$,
\begingroup
\allowdisplaybreaks
\begin{align}
S =
& 
\int dt \int d^{2} \boldsymbol{x} \int dz \,
\Big[
- \frac{ \epsilon_{0} c^{2} }{4}
F_{\mu\nu} F^{\mu\nu}
+ \frac{i \hbar }{2} (\bar{\Psi} \mathcal{D}_{t} \Psi 
- \overline{\mathcal{D}_{t} \Psi} \Psi ) 
- \frac{\hbar^{2}}{2m} \overline{ \mathcal{D}_{i}\Psi}\mathcal{D}_{i}\Psi
\nonumber\\
&~~~~~~~~~~~~~~~~~~~~~~~~~
+ \frac{1}{2v_{N}^{2}}(\partial_{t}N)^{2}
-\frac{1}{2}(\partial_{i}N)^{2}
-V(|\Psi|,N) + q n_{\rm s} \Phi
\Big]
,
\label{201}
\end{align}
\endgroup
where the field strength tensor $F_{\mu\nu}=\partial_{\mu}A_{\nu}-\partial_{\nu}A_{\mu}$ is related to electric and magnetic fields, $(\boldsymbol{E})^{i} = c F_{i0}$ and $\displaystyle (\boldsymbol{B})^{i} = \frac{1}{2} \epsilon^{ijk} F_{jk}$, a Schr\"{o}dinger type matter $\Psi=|\Psi|e^{i\Omega}$ couples minimally to the U(1) gauge field,
\begin{align}
\mathcal{D}_{t} \Psi
= \Big( \frac{\partial}{\partial t} + i \frac{q}{ \hbar} \Phi \Big) \Psi
,\qquad
\mathcal{D}_{i} \Psi
= \Big( \frac{\partial}{\partial x^{i}} - i \frac{q}{\hbar} A^{i} \Big) \Psi
,\label{203}
\end{align}
and the constant matter density of charged background called the superfluid density $n_{\rm s}$ couples electrically to the scalar potential $\Phi$.
The potential of neutral and complex scalar fields $V$ involves quartic self-interaction of nonnegative coupling $\lambda$ and cubic Yukawa type interaction between neutral and complex scalar fields of nonpositive coupling $g$,  
\begin{align}
V(|\Psi|,N) = \lambda (|\Psi|^{2} - v^{2}) \Big( |\Psi|^{2} + \frac{g}{\lambda} N - v^{2} \Big)
,
\label{206}
\end{align}
where $v$ is vacuum expectation value of the complex scalar field amplitude $|\Psi|$.
The gapless neutral scalar field of acoustic phonon $N$ represents a collection of residual low energy modes lower than the UV cutoff given by the Debye frequency $\omega_{{\rm D}}\approx1.38\times10^{13\sim14}{\rm Hz}$.
High frequency modes of the phonon are integrated out and generate the quartic self-interaction which lets effective field theory of the effective action \eqref{201} renormalizable.

Physically reliable classical vacuum is achieved as the zero energy configuration of specific constant matter fields,
\begin{align}
(\braket{|\Psi|}, \braket{N}) = (v,0) ,
\label{221}
\end{align}
with zero electric field, $\boldsymbol{E}={\bf 0}$, explaining perfect conductivity and zero magnetic field,
$\boldsymbol{B}={\bf 0} $. The obtained vacuum is stable along the scalar amplitude and is flat along the neutral scalar field.
Nonzero vacuum expectation value $\braket{|\Psi|} = v$ called the rigidity of wavefunction in superconductivity gives spontaneous breakdown of the $\text{U}(1)$ symmetry and electrical neutrality of this Higgs vacuum leads to the identification of the vacuum expectation value $v$ in terms of the superfluid density $n_{\rm s}$,
\begin{align}
v^{2} = n_{\rm s}
.\label{305}
\end{align} 
Even in the presence of external constant uniform magnetic field, $\boldsymbol{B}^{\rm ext}=(0,0,B^{\rm ext})$, the recovery of zero energy leads to cancellation of the external magnetic field whose gauge field under the symmetric gauge is $\displaystyle A^{i} = -\frac{1}{2} \epsilon^{ij} x_{j} B^{\rm ext}$.
Therefore, the Meissner effect of perfect diamagnetism is understood as the recovery of the superconducting vacuum of zero energy in response to the external magnetic field.
Note that all other components of the energy-momentum tensor $T^{\mu}_{~\nu}$, energy flux, momentum density, and stress components, vanish in the symmetry-broken vacuum \eqref{221}.

In the framework of perturbative quantum field theory, attractive and repulsive static forces between two Cooper pairs are cancelled in superconducting samples with the critical couplings.
In the tree level, the repulsive electrostatic Coulomb force mediated by the massless degree of $\mathrm{U}(1)$ gauge field is perfectly cancelled by the attractive force mediated by the gapless acoustic phonon in its static limit for the critical coupling of cubic Yukawa type interaction between phonon and Cooper pair,
\begin{align}
g_{\rm c} &= \frac{q}{\sqrt{\epsilon_{0}}}
= - 1.08\times10^{-13} \, {\rm kg}^{1/2} {\rm m}^{3/2} {\rm s}^{-1}
.\label{327}
\end{align}
Existence of the critical coupling of cubic Yukawa type interaction between phonon and Cooper pair explains the reason why two negatively charged Cooper pairs behave noninteracting despite of the Coulomb repulsion in superconducting phase and predicts a criterion for conventional superconductivity in this effective field theory.
Therefore, superconducting state can possibly be materialized when the phonon coupling is close to its critical value.
In addition, low critical temperature and temperature-fragile character of conventional superconductivity seem to be originated from huge discrepancy between the two propagation speeds of light and phonon of the order $v_{N}/c \sim 10^{-5}$.

In the superconducting phase, there are two characteristic length scales:
One is the correlation length for the Higgs field,
\begin{align}
\xi =
\frac{\hbar}{2} \frac{1}{\sqrt{2 m\lambda n_{\rm s}}}
,\label{303}
\end{align}
whose value of a Cooper pair is $\xi = 2.76 \times 10^{14} / \sqrt{\lambda n_{\rm s}} ~ \text{kg}^{\frac{1}{2}}\, \text{m}^{2}\, \text{s}^{-1}$.
The other is the London penetration depth for the massive gauge field,
\begin{align}
\lambda_{{\rm L}} = \sqrt{\frac{\epsilon_{0} m}{n_{\rm s}}} \frac{c}{|q|}
,\label{302}
\end{align}
whose value of the Cooper pair is $\lambda_{\rm L} = 3.76 \times 10^{6} / \sqrt{n_{\rm s}} ~\text{m}^{-\frac{1}{2}}$.
Thus their competition is expressed by the ratio called the Ginzburg-Landau parameter
\begin{align}
\kappa = \frac{\lambda_{\rm L}}{\xi} 
= \frac{2mc \sqrt{2\epsilon_{0} \lambda}}{\hbar |q|} 
.\label{349}
\end{align}

When the two length scales \eqref{303}--\eqref{302} are equal,
the Ginzburg-Landau parameter becomes unity\footnote{%
$\sqrt{2}$ is used in condensed matter community rather than unity.
Since 2 in the $\sqrt{2}$ is originated from $1/2$ in front of quadratic spatial derivative term in the nonrelativistic linearized scalar equation, we use the definition of correlation length \eqref{303} divided by $\sqrt{2}$ throughout this paper.%
}%
\begin{align}
\kappa = \frac{\lambda_{\rm L}}{\xi} = 1
,\label{314}
\end{align}  
and the quartic self-interaction of complex scalar field has the critical coupling for a Cooper pair,
\begin{align}
\lambda_{\rm c} = \frac{\hbar^{2} q^{2}}{8\epsilon_{0} m^{2} c^{2}} 
= 2.14 \times10^{-52} ~\text{kg}\, \text{m}^{5}\, \text{s}^{-2}
.\label{312}
\end{align}
The known critical coupling $\lambda_{\rm c}$ \eqref{312} is understood as another 1-loop level interaction balance between the short ranged electromagnetic repulsion mediated by massive degree of the $\mathrm{U}(1)$ gauge field and the short ranged attraction mediated by the massive neutral Higgs boson, in its static limit.
According to the critical coupling of quartic self-interaction of complex scalar field \eqref{312}, superconducting materials are classified into two types.
In weak coupling regime $0<\lambda<\lambda_{{\rm c}}$, the net interaction becomes repulsive in long distance that leads to type I superconductivity and, in strong coupling regime $\lambda>\lambda_{{\rm c}}$, attractive in long distance that leads to type I$\!$I as predicted firstly in the Ginzburg-Landau theory\cite{Abrikosov:1956sx}.

Thus these two critical couplings can be understood perturbatively through the two balances of attractive and repulsive static forces including quantum corrections, and now noninteracting nature of the Cooper pairs in superconducting phase despite of their electric charges is reconciled in the context of effective field theory.
In synthesis, superconducting materials predict manifestation of their superconducting phase only at or in the vicinity of the critical phonon coupling \eqref{327} and in the static limit of interactions.
Conventional superconductors of $s$-wave are classified into two types I and I$\!$I according to the competition of net short ranged repulsion and attraction in the framework of the proposed effective field theory with the action \eqref{201}.

\section{Borderline of Types of Superconductivity as BPS Limit}

Effective field theory of the action \eqref{201} of a complex scalar field of $s$-wave is proposed and perturbative analysis shows that superconducting materials realize their superconducting phase at the critical phonon coupling \eqref{327} and are classified into two types whose borderline is given by the critical coupling of $\lambda=\lambda_{{\rm c}}$ \eqref{312}.
In this section, we reexamine the effective field theory of critical couplings by use of nonperturbative analysis and show in detail that the critical point $(\lambda_{{\rm c}},g_{{\rm c}})$ is identified as the conditions to achieve the Bogomolny-Prasad-Sommerfield (BPS) limit for topological vortices, the representative nonperturbative spectra supported in type I$\!$I superconductors \cite{FTSC}.

Suppose that superconducting samples take the shape of thin or thick slab which is usually invariant along the direction of the slab thickness denoted by $z$ axis.
This translation symmetry along \(z\) axis dictates independence of $z$ coordinate of the fields,
$
\partial_{3} \Psi
=\partial_{3}\Phi= \partial_{3} A^{i}
=\partial_{3}N= 0
$,
and, consistently, the \(z\) component of gauge field is assumed to be turned off,
$
A^{3} = 0
$.
Accordingly, some components of covariant derivative and electric and magnetic fields vanish,
$ \mathcal{D}_{3} \Psi = 0$ and
$
(\boldsymbol{E})^{3}=(\boldsymbol{B})^{1}
= (\boldsymbol{B})^{2}
= 0
$.
Then the action \eqref{201} in $(1+3)$ dimensions becomes independent of $z$ coordinate and the sample, e.g., a superconducting slab, can be regarded theoretically as a planar system described by the following $(1+2)$-dimensional action
\begingroup
\allowdisplaybreaks
\begin{align}
\bar{S}
=&\, \frac{S}{\int dz} \nonumber\\
=&\, 
\int dt \int d^{2} \boldsymbol{x} \,
\Big[ \frac{\epsilon_{0}}{2} (\boldsymbol{E}^{2} - c^{2} B^{2}) + q n_{\rm s} \Phi
+ \frac{i \hbar }{2} (\bar{\Psi} \mathcal{D}_{t} \Psi 
- \overline{\mathcal{D}_{t} \Psi} \Psi ) 
-\frac{\hbar^{2}}{2m} \overline{ \mathcal{D}_{i}\Psi}\mathcal{D}_{i}\Psi
\nonumber\\
&~~~~~~~~~~~~~~~~~~
+ \frac{1}{2v_{N}^{2}}(\partial_{t}N)^{2}
-\frac{1}{2}(\partial_{i}N)^{2}
-V(|\Psi|,N)
\Big]
,
\label{301}
\end{align}
\endgroup
where the electric field has two planar components $\boldsymbol{E} = (E^{1} , E^{2})$, the $z$ component of magnetic field is set as 
\begin{align}
(\boldsymbol{B})^{3} = B^{3} = B
=\partial_{1}A^{2}-\partial_{2}A^{1}
,\label{313}
\end{align}  
and the index $i$ runs $1$ and $2$ such as
$\displaystyle
\boldsymbol{\nabla}
= (\partial_{1}, \partial_{2})
= \Big( \frac{\partial}{\partial x}, \frac{\partial}{\partial y} \Big)
$
in what follows. 
Energy per unit length along $z$ axis for static field configurations symmetric under the translation along $z$ axis is read from the action \eqref{301},
\begingroup
\allowdisplaybreaks
\begin{align}
\bar{E}
& = \frac{E}{\int dz} \nonumber \\
& = \int d^{2} \boldsymbol{x} \,
\bigg\{
\boldsymbol{\nabla} \cdot ( \epsilon_{0} \Phi \boldsymbol{E} )
- \epsilon_{0} \Phi \Big[ \boldsymbol{\nabla} \cdot \boldsymbol{E} 
- 
\frac{q}{\epsilon_{0}} (|\Psi|^{2} - n_{\rm s}) \Big]
\nonumber\\
&~~~~~~~~~~~~~~~
+
\frac{\epsilon_{0}}{2} (\boldsymbol{E}^{2} + c^{2}B^{2} ) 
+ \frac{\hbar^{2}}{2m}(\boldsymbol{\nabla} |\Psi|)^{2}
+ \frac{q^{2}}{2m} |\Psi|^{2}\Big( A^{i} - \frac{\hbar}{q} \partial_{i} \Omega \Big)^{2}
+ \frac{1}{2} (\boldsymbol{\nabla} N)^{2}
\nonumber\\
&~~~~~~~~~~~~~~~
+ \lambda (|\Psi|^{2} - v^{2}) \Big( |\Psi|^{2} + \frac{g}{\lambda} N - v^{2} \Big)
\bigg\}
.
\label{361}
\end{align}
\endgroup 

The set of the superselective symmetry-broken vacuum configurations constitutes a cylinder ${\rm S}^{1} \times \mathbb{R}^{1}$ of radius $v$ as the vacuum manifold whose topology is characterized by the winding number $n$ 
($n\in\mathbb{Z}$) called the vorticity.
The obtained Higgs vacuum \eqref{221} also called the perfect superconducting state corresponds to topologically trivial sector of zero winding number $n=0$.
From now on we take into account topologically nontrivial sector by turning on nonzero winding number $n$.
In the symmetry-broken phase of 
$\displaystyle{\lim_{|\boldsymbol{x}|\rightarrow\infty}|\Psi|
=\sqrt{n_{{\rm s}}}}$, there can exist topologically stable vortex solutions whose complex scalar field behaves along the circle at spatial infinity $ \partial \mathbb{R}^{2}_{(x, y)} = \mathrm{S}^{1} $ in such a manner that
\begin{align}
\oint_{ |\boldsymbol{x}| \to \infty } d \boldsymbol{l} \cdot \nabla \ln |\Psi|
= 2\pi i n
.
\end{align}
As we shall see, the topological solitons supported in topologically nontrivial sectors of the $\mathrm{U}(1)$ gauge theory of our consideration are the gauged vortices classified by the magnetic flux $\Phi_{B}$ called mathematically the first Chern character.
Since the magnetic field of interest is parallel to $z$ axis, $\boldsymbol{B} = (0,0,B)$ \eqref{313}, and depends only on planar coordinates, $B = B(x,y)$, the magnetic flux defined on the $xy$ plane is computed with the help of the vanishing spatial components of covariant derivative \eqref{203} at spatial infinity,
\begin{align}
\Phi_{B} = \int_{\mathbb{R}^{2}} d^{2} \boldsymbol{x} \, B 
= \oint_{\partial \mathbb{R}^{2}} d \boldsymbol{l} \cdot \boldsymbol{A}
=- 2\pi\Phi_{\rm L} n  
,\label{402}
\end{align}
where the nonzero vector potential is also chosen to be a planar vector $\boldsymbol{A} = (A^{1} , A^{2})$ and $\Phi_{{\rm L}}$ is the London flux quantum
\begin{align}
\Phi_{{\rm L}} = \frac{\hbar}{q}
.
\end{align}

When the quartic self-interaction coupling has the critical value \eqref{312}, the correlation length \eqref{303} and the London penetration depth \eqref{302} become equal and this limit of unit Ginzburg-Landau parameter \eqref{314} defines the borderline of type I and I$\!$I superconductors~\cite{Abrikosov:1956sx}, which is perturbatively understood as an interaction balance in static limit \cite{NRAH}.
Another characteristic property in the critical couplings in \eqref{327} and \eqref{312} appears in the energy $\bar{E}$ per unit length along $z$ axis \eqref{361}, which is reorganized by use of the Bogomolny trick \cite{Bogomolny:1975de}, 
\begingroup
\allowdisplaybreaks
\begin{align}
\bar{E} =&\, \int d^{2} \boldsymbol{x}\, \bigg\{  
- \sqrt{\epsilon_{0}} (N + \sqrt{\epsilon_{0}} \Phi) 
\Big[ \boldsymbol{\nabla}\cdot\boldsymbol{E} 
- \frac{q}{\epsilon_{0}} (|\Psi|^{2} - n_{\rm s}) \Big]  
+ \nabla\cdot \big[ \sqrt{\epsilon_{0}} \boldsymbol{E} (N + \sqrt{\epsilon_{0}} \Phi) \big] 
\nonumber\\
&\,~~~~~~~~~~ + \frac{1}{2} (\partial_{i} N - \sqrt{\epsilon_{0}} E^{i})^{2} + \frac{\hbar^{2}}{2m} |(\mathcal{D}_{1} \mp i \mathcal{D}_{2}) \Psi|^{2} + \frac{\epsilon_{0} c^{2}}{2} \bigg[ B \mp \frac{\hbar}{2q\lambda_{\rm L}^{2}} \bigg( \frac{|\Psi|^{2}}{n_{\rm s}} - 1 \bigg) \bigg]^{2}  \nonumber\\
&\,~~~~~~~~~~
\pm \partial_{i} (\epsilon^{ij} j_{j})\mp \frac{\hbar q}{2m} n_{\rm s} B \bigg\}
,\label{424}
\end{align}
\endgroup
where $j_{j} = j^{j}$ is the U(1) current density
\begin{equation}
(\boldsymbol{j})^{i}
= j^{i}
=- i \frac{q \hbar}{2m} \big(
\bar{\Psi} \mathcal{D}_{i} \Psi
- \overline{\mathcal{D}_{i} \Psi} \Psi
\big)
,
\label{202}
\end{equation} and the commutation relation
$\displaystyle
[\mathcal{D}_{1}, \mathcal{D}_{2}]
= - i \frac{q}{\hbar} F_{12} = - i \frac{q}{\hbar} B
$
is used.
The first term in the first line vanishes with the help of Gauss' law
\begin{align}
\boldsymbol{\nabla} \cdot \boldsymbol{E} = \frac{\rho}{\epsilon_0}
,\label{209}
\end{align}
where charge density is
\begin{align}
\rho = q(|\Psi|^{2} - n_{\rm s}) 
,
\label{215}
\end{align}
and the N\"{o}ther charge is identified with the electric charge
\begin{align}
Q_{{\rm U}(1)} = \int d^{2}\boldsymbol{x}dz\,\rho
.\label{228}
\end{align}
Since the theory possesses translation symmetry along $z$ axis, physically relevant quantity is not the electric charge $Q_{\mathrm{U}(1)}$ \eqref{228} but charge per unit length along $z$ axis,
\begin{align}
\bar{Q}_{{\rm U(1)}}
= \frac{ Q_{{\rm U(1)}} }{\int dz}
= q \int d^{2} \boldsymbol{x} \,
(|\Psi|^{2} - n_{\rm s})
.\label{315}
\end{align}If the constant background charge density $q n_{\rm s}$ is not taken into account, its coupling to the scalar potential $qn_{{\rm s}}\Phi$ is missing in the action \eqref{201} and the charge density becomes $\rho = q |\Psi|^{2}$ \cite{Wen:2004ym}.
Then finiteness of the energy and the $\text{U}(1)$ charge are not achieved simultaneously in classical level, e.g., those of the vortices.
Moreover, the term of current density $j_{j}$ does not contribute to the energy $\bar{E}$ for the static vortices of our consideration.

If the following Bogomolny equations hold, 
\begingroup
\allowdisplaybreaks
\begin{align}
\boldsymbol{\nabla} N - \sqrt{\epsilon_{0}} \boldsymbol{E}
&=0
,\label{422}
\\
(\mathcal{D}_{1} \mp i \mathcal{D}_{2}) \Psi
&=0
,\label{425}
\\
B \mp \frac{\hbar}{2q\lambda_{\rm L}^{2}} \bigg( \frac{|\Psi|^{2}}{n_{\rm s}} - 1 \bigg)
&=0
,\label{427}
\end{align}
\endgroup
the energy bound so-called the BPS bound~\cite{Bogomolny:1975de} is saturated in terms of the magnetic flux \eqref{402} whose quantized value is proportional to vorticity $n$,
\begingroup
\allowdisplaybreaks
\begin{align}
\bar{E} 
\ge \frac{\hbar n_{\rm s}}{2m} |q\Phi_{B}|
= \frac{\hbar^{2} n_{\rm s}}{2m}
\bigg| \frac{\Phi_{B}}{\Phi_{\rm L}} \bigg|
= \frac{\pi \hbar^{2} n_{\rm s} }{m} |n|
= \frac{\hbar^{2} n_{\rm s}}{4 \epsilon_{0} m^{2} c^{2}}
| q \bar{Q}_{\mathrm{U}(1)} |
,\label{421}
\end{align}
\endgroup
 where the equality holds for either $n$ vortices or $n$ antivortices. In the BPS limit of critical coupling $\lambda=\lambda_{\rm c}$ \eqref{312}, the electric charge per unit length along $z$ axis \eqref{315} can be computed exactly by comparing the Bogomolny equation \eqref{427} and charge density $\rho$ \eqref{215},
\begin{align}
\bar{Q}_{\text{U}(1)}
= \frac{2\epsilon_{0} m c^{2}}{\hbar} |\Phi_{B}|
= - \frac{4\pi\epsilon_{0} m c^{2}}{q} |n|
,\label{434}
\end{align}
which is also quantized by vorticity $n$. Thus the last equality in \eqref{421} also holds.
It is straightforward to check that every static solution of the Bogomolny equations \eqref{422}--\eqref{427} automatically satisfies the second-order Euler-Lagrange equations.
Note that inverse proportionality between the vortex charge and the Cooper pair charge, $ Q_{\mathrm{U}(1)} \propto 1/q $, implies a duality between strong and weak couplings.
Subsequently all the stress components $T^{i}_{~j}$ of the canonical energy momentum tensor $T^{\mu}_{~\nu}$ vanish everywhere for any BPS configurations, specifically for both BPS vortices and antivortices,
\begingroup
\allowdisplaybreaks
\begin{align}
T^{i}_{~j} =&\, -\frac{1}{2} (\partial_{k} N - \sqrt{\epsilon_{0}} E^{k}) (\partial_{k} N + \sqrt{\epsilon_{0}} E^{k}) \delta_{ij} 
+ (\partial_{i} N - \sqrt{\epsilon_{0}} E^{i}) (\partial_{j} N + \sqrt{\epsilon_{0}} E^{j})
\nonumber\\
&\, + \frac{\epsilon_{0} c^{2}}{2} \bigg[ B - \frac{\Phi_{\rm L}}{2\lambda_{\rm L}^{2}} \bigg( 1 - \frac{|\Psi|^{2}}{n_{\rm s}} \bigg) \bigg] \bigg[ B + \frac{\Phi_{\rm L}}{2\lambda_{\rm L}^{2}} \bigg( 1 - \frac{|\Psi|^{2}}{n_{\rm s}} \bigg) \bigg] \delta_{ij} \nonumber\\
&\, + \frac{\hbar^{2}}{8m} \big[ \overline{(\mathcal{D}_{i} - i \epsilon^{ik} \mathcal{D}_{k}) \Psi} (\mathcal{D}_{j} + i \epsilon^{jl} \mathcal{D}_{l}) \Psi + \overline{(\mathcal{D}_{i} + i \epsilon^{ik} \mathcal{D}_{k}) \Psi} (\mathcal{D}_{j} - i \epsilon^{jl} \mathcal{D}_{l}) \Psi \nonumber\\
&\, ~~~~~~~~~ + \overline{(\mathcal{D}_{j} - i \epsilon^{jk} \mathcal{D}_{k}) \Psi} (\mathcal{D}_{i} + i \epsilon^{il} \mathcal{D}_{l}) \Psi + \overline{(\mathcal{D}_{j} + i \epsilon^{jk} \mathcal{D}_{k}) \Psi} (\mathcal{D}_{i} - i \epsilon^{il} \mathcal{D}_{l}) \Psi \big] 
\nonumber\\
&\, - \sqrt{\epsilon_{0}} \big[ \partial_{i} (N E^{j}) - \partial_{j} (N E^{i}) \big] - \frac{q}{\sqrt{\epsilon_{0}}} (N + \sqrt{\epsilon_{0}} \Phi) (|\Psi|^{2} - n_{\rm s}) \delta_{ij}
\nonumber\\
=
&\,
0.\label{420}
\end{align}
\endgroup  
The two consequences, the energy per unit length along $z$ axis exactly proportional to the number of vortices or antivortices \eqref{421} and no interaction at every spatial point \eqref{420}, lead to a conclusion that BPS vortices or BPS antivortices are noninteracting at least in the classical level by disappearance of any net interaction due to exact cancellation of attractive and repulsive interactions.
This nonperturbative result, the BPS limit of static noninteracting vortices, in the current section, matches the perturbative result, the static force balance between two Cooper pairs, in the previous section. 

The first Bogomolny equation \eqref{422} is solved to relate the static neutral scalar field to the static scalar potential as
\begin{align}
\Phi = - \frac{1}{\sqrt{\epsilon_{0}}} N
, \label{423}
\end{align}
where the redundant constant gauge degree of freedom is fixed to be zero. The obtained relation for static vortex solutions in the BPS limit shows nonperturbative cancellation of the electrostatic force by the static force mediated by the neutral scalar field.
The second Bogomolny equation \eqref{425} expresses spatial components of the $\text{U}(1)$ gauge field by the complex scalar amplitude and phase
\begin{align}
A^{i} = \frac{\hbar}{q} (\mp \epsilon^{ij} \partial_{j} \ln |\Psi| + \partial_{i} \Omega)
.\label{426}
\end{align}
Substitution of the gauge field \eqref{426} into the third Bogomolny equation \eqref{427} results in a second order elliptic equation for the scalar amplitude
\begin{align}
\nabla^{2} \ln \frac{|\Psi|^{2}}{\displaystyle \prod_{a=1}^{n} |\boldsymbol{x} - \boldsymbol{x}_{a}|^{2}} = \frac{1}{\lambda_{\rm L}^{2}} \bigg( \frac{|\Psi|^{2}}{n_{\rm s}} - 1 \bigg)
.\label{428}
\end{align}
The resultant Bogomolny equation \eqref{428} is nothing but that of the Abelian Higgs model and the Ginzburg-Landau theory, and supports static topological multi-vortex solutions of arbitrary separations.
More precisely, existence and uniqueness of $n$ separated BPS vortex solution satisfying the boundary conditions,
\begin{align}
\lim_{\boldsymbol{x} \to \boldsymbol{x}_{a}} |\Psi| (\boldsymbol{x}) = 0 \quad \text{and} \quad \lim_{|\boldsymbol{x}|\to\infty} |\Psi| (\boldsymbol{x}) = \sqrt{n_{\rm s}} \,
, \label{429}
\end{align}
are proved with mathematical rigor \cite{Taubes:1979tm}, in which each arbitrary position $\boldsymbol{x}_{a} = (x_{a}, y_{a})~(a=1,2,\cdots,n)$ stands for $a$-th vortex site.
These $2n$ planar position coordinates $(x_{a},y_{a})$ of the $n$ separated vortices of arbitrary interdistances are identified as the $2n$ zero modes calculated by use of the Atiyah-Singer index theorem \cite{Weinberg:1979er}.
Existence of the $2n$ zero modes of zero energy cost is consistent with 
noninteracting topological vortices independent of the interdistances in the BPS limit of the critical quartic self-interaction coupling $\lambda = \lambda_{{\rm c}}$ \eqref{312}. 
Note that a boundary condition at spatial infinity in \eqref{429} introduced for finite energy is consistent with localization of the U(1) charge density \eqref{215}.

In summary, we show that the BPS bound \eqref{421} is saturated for static vortex configurations only at the critical couplings of quartic self-interaction of Cooper pairs and cubic Yukawa type interaction between phonon and Cooper pairs, $(\lambda, g) = (\lambda_{\rm c}, g_{\rm c})$ as in \eqref{312} and \eqref{327}.
In the previous section, this limit at the critical couplings is explained in perturbative regime by the static force balances achieved through perfect cancellation of attraction and repulsion.
Therefore, in the current section, those balances are confirmed in nonperturbative regime by the BPS limit of noninteracting multi-vortices.

\section{Charged Vortex}

In this section, we consider the effective field theory of the action \eqref{201} of couplings $\lambda$ $(\lambda\ge0)$ and $g$ $(g\le0)$ and investigate nonperturbative solitonic configurations of nonzero energy in the $xy$ plane with translation symmetry along $z$ axis.
Specifically the characters of topological vortices and their interactions are studied with application to superconductors.

Since constant scalar phase is periodic, $ \Omega \in \mathbb{R} / 2\pi \mathbb{Z} $, and each vacuum $ (|\Psi|, \Omega) = (v, \Omega) $ is superselective for every sample with macroscopic volume, manifold of the symmetry-broken vacua is a cylinder $\text{S}^{1} \times \mathbb{R}^{1}$ of radius $|\Psi| = v$ with the help of the chosen zero neutral scalar field \eqref{221} and the spatial infinity $|\boldsymbol{x}| \to \infty$ forms another circle $\partial \mathbb{R}^{2}_{ (x, y) } = \text{S}^{1}$ of infinite radius.
Hence the winding between the cylinder and the circle is classified by the first homotopy, $\Pi_{1} ( \text{S}^{1} \times \mathbb{R}^{1} ) = \mathbb{Z}$.
In the Ginzburg-Landau theory and the Abelian Higgs model, topologically stable gauged vortices of finite energy, called the Abrikosov-Nielsen-Olesen vortices, are supported \cite{Abrikosov:1956sx, Nielsen:1973cs}.
We study topological vortices in the proposed effective field theory and discuss application mostly to conventional superconductivity \cite{Bennemann:2008, tinkham2004introduction, Arovas:2019, cohen2016fundamentals}.

 The energy per unit length along $z$ axis \eqref{361} for static field 
configurations symmetric under the translation along $z$ axis is rewritten as
\begingroup
\allowdisplaybreaks
\begin{align}
\bar{E}& = \int d^{2} \boldsymbol{x} \,
\bigg\{
\frac{\epsilon_{0} c^{2}}{2} B^{2}
+ \frac{\hbar^{2}}{2m} (\boldsymbol{\nabla} |\Psi|)^{2}
+ \frac{q^{2}}{2m} |\Psi|^{2}\Big( A^{i} - \frac{\hbar}{q} \partial_{i} \Omega \Big)^{2}
+ \lambda (|\Psi|^{2} - v^{2})^{2}
\nonumber\\
&~~~~~~~~~~~~~~~
- \frac{\epsilon_{0}}{2} \Big[ \boldsymbol{E}^{2} 
- \frac{q^{2}}{\epsilon_{0}^{2} g^{2}} (\boldsymbol{\nabla} N)^{2} \Big]
+q \Big( \Phi + \frac{q}{\epsilon_{0} g}N \Big)(|\Psi|^{2} - n_{\rm s}) 
\nonumber\\
&~~~~~~~~~~~~~~~ 
+\Big( 1 -\frac{q^{2}}{\epsilon_{0}^{2} g^{2}}  \Big)\Big[
\frac{1}{2} 
 (\boldsymbol{\nabla} N)^{2}+  gN(|\Psi|^{2} - n_{\rm s})
\Big]
\bigg\}
.
\label{460}
\end{align} 
\endgroup 
With arbitrary couplings $\lambda$ and $g$, classical dynamics is governed by the Euler-Lagrange equations.
The charged complex scalar field follows the gauged nonlinear Schr\"{o}dinger equation
\begin{equation}
i \hbar \mathcal{D}_{t}\Psi
=
- \frac{\hbar^{2}}{2m} \mathcal{D}_{i}^{2}\Psi
-
2 \lambda (v^{2}- |\Psi|^{2} )\Psi+ g N  \Psi
.\label{213}
\end{equation}
The Maxwell equations for the U(1) gauge field \(A^\mu\) consist of the Gauss' law \eqref{209} with the charge density \eqref{215} and the Amp\'{e}re's law including displacement current term with the current density \eqref{202},
\begin{align}
- \frac{1}{c^{2}} \frac{\partial \boldsymbol{E}}{\partial t}
+ \boldsymbol{\nabla} \times \boldsymbol{B} 
= \frac{\boldsymbol{j}}{\epsilon_{0} c^{2}}
\label{222}
.
\end{align}
The neutral scalar field equation is given by an inhomogeneous linear wave equation for acoustic waves,
\begin{align}
\frac{1}{v_{N}^{2}}\frac{\partial^{2}N}{\partial t^{2}}-\nabla^{2}N
= - g (|\Psi|^{2} - v^{2})
.\label{214}
\end{align}
Once a solution of static scalar amplitude $|\Psi|$ is achieved, it is exactly solved for static neutral scalar field in three and two spatial dimensions,
\begin{align}
N (\boldsymbol{x}, z)
= - \frac{g}{4 \pi}
\int d^{2} \boldsymbol{x}^{\prime} d z^{\prime}
\frac{ |\Psi|^{2} ( \boldsymbol{x}^{\prime} , z^{\prime} ) - v^{2} }
{ \sqrt{ |\boldsymbol{x}^{\prime} - \boldsymbol{x}|^{2} + |z^{\prime} - z|^{2} } }
,\label{440}
\end{align}
\begin{align}
N (\boldsymbol{x})
= \frac{g}{2 \pi}
\int d^{2} \boldsymbol{x}^{\prime} \,
[ |\Psi|^{2} ( \boldsymbol{x}^{\prime} ) - v^{2} ]
\ln |\boldsymbol{x}^{\prime} - \boldsymbol{x}|
.\label{473}
\end{align}
Since the vacuum expectation value is identified as square root of the constant  superfluid density \eqref{305}, the right-hand side of the Gauss' law \eqref{209} becomes identical as that of the neutral scalar equation \eqref{214} up to a proportionality constant.
For time-independent configurations, the scalar potential is linearly related to the neutral scalar field,
\begin{align}
\Phi = - \frac{q}{\epsilon_{0}g} N + f
,\label{310}
\end{align}
where $f=f(\boldsymbol{x})$ is an arbitrary function whose Laplacian vanishes, $\nabla^{2} f=0$, and is nothing but the residual gauge degree of freedom under the Coulomb gauge, which can always be fixed by zero, $f=0$.
Hence the relation \eqref{310} with the trivially fixed redundant gauge degree of freedom $f=0$  is equivalent to  \eqref{423} from the Bogomolny equation \eqref{422} only if the cubic Yukawa type coupling $g$ has the critical value $g_{\rm c}$ \eqref{327} in the BPS limit.

For simplicity, we deal mostly with static cylindrically symmetric vortex solutions of vorticity $n~(0 \neq n \in \mathbb{Z})$ by examining the coupled nonlinear equations.
With trivial form of $\Phi=\Phi(r)$ and $N=N(r)$, the ansatz compatible with cylindrically symmetric vortex configurations is
\begingroup
\allowdisplaybreaks
\begin{align}
\Psi =&\, |\Psi| e^{-i n\theta}, \label{413}\\
A^{i} =&\, - \frac{\hbar}{q} \epsilon^{ij} x_{j} \frac{A}{r^{2}} ,\label{414}
\end{align}
\endgroup
where $r=\sqrt{x^{2}+y^{2}}$ and $\theta=\tan^{-1}(y/x)$. 
For cylindrically symmetric configurations \eqref{413}--\eqref{414} including vortices, the energy per unit length along $z$ axis becomes
\begingroup
\allowdisplaybreaks
\begin{align}
\bar{E} &
= 2\pi \int_{0}^{\infty} dr \, r \,
\bigg\{
\frac{\epsilon_{0} c^{2}\hbar^{2}}{2q^{2}}\frac{1}{r^{2}}
\Big(  \frac{dA}{dr} \Big)^{2}
+ \frac{\hbar^{2}}{2m} \Big( \frac{d|\Psi|}{dr} \Big)^{2}
+ \frac{\hbar^{2}}{2m} \frac{(A+n)^{2}}{r^{2}}  |\Psi|^{2}
+ \lambda (|\Psi|^{2} - v^{2})^{2}
\nonumber \\
& \hspace{7em}
- \frac{\epsilon_{0}}{2}
\Big[ \Big( \frac{d\Phi}{dr} \Big)^{2} - \frac{q^{2}}{\epsilon_{0}^{2} g^{2}} \Big( \frac{dN}{dr} \Big)^{2} \Big]
+q \Big( \Phi + \frac{q}{\epsilon_{0} g} N \Big)(|\Psi|^{2} - n_{\rm s} )
\nonumber \\
& \hspace{7em}
+\Big( 1 -\frac{q^{2}}{\epsilon_{0}^{2} g^{2}}  \Big)\Big[\frac{1}{2} 
 \Big( \frac{dN}{dr} \Big)^{2}  
 + g N(|\Psi|^{2} - n_{\rm s} )
\Big]
\bigg\} 
.
\label{417}
\end{align} 
\endgroup
  The equation of complex scalar field \eqref{213} and the Amp\'{e}re's law \eqref{222} become,
\begingroup
\allowdisplaybreaks
\begin{align}
\frac{d^{2} |\Psi|}{dr^{2}} + \frac{1}{r} \frac{d|\Psi|}{dr} 
& = \frac{(A+n)^{2}}{r^{2}} |\Psi| 
+ \frac{1}{2\xi^{2}} \Big(\frac{|\Psi|^{2}}{n_{\rm s}} - 1\Big) |\Psi| + \frac{1}{4\lambda n_{\rm s} \xi^{2}} (g N + q \Phi) |\Psi|
,
\label{470}
\\
\frac{d^{2} A}{dr^{2}} - \frac{1}{r} \frac{dA}{dr} 
& = \frac{A+n}{\lambda_{\rm L}^{2}} \frac{|\Psi|^{2}}{n_{\rm s}} 
.
\label{418}
\end{align}
\endgroup
For a given static cylindrically symmetric complex scalar amplitude $|\Psi|$, both scalar potential $\Phi$  and neutral scalar field $N$ are obtained by 
\begingroup
\allowdisplaybreaks
\begin{align}
\Phi(r)= - \frac{q}{\epsilon_{0}g} N(r)
=- \frac{q}{4 \pi\epsilon_{0}} 
\int_{0}^{\infty} dr^{\prime}r^{\prime} [|\Psi|^{2} (r^{\prime}) - n_{{\rm s}}]
\int_{0}^{2\pi}d\theta \,
\ln |r^{\prime 2}+r^{2}-2r^{\prime}r\cos\theta|
.
\label{471}
\end{align} 
\endgroup
Necessary conditions for regular behavior of the vortex solution and finiteness of the energy per unit length along $z$ axis \eqref{417} determine the appropriate boundary conditions at the origin and at spatial infinity. Scalar amplitude and vector potential obey, 
\begingroup
\allowdisplaybreaks
\begin{align}
&~|\Psi| (0) = 0, \qquad\qquad\qquad A (0) = 0
,
\label{415}
\\
&\lim_{r\to\infty} |\Psi| = v = \sqrt{n_{\rm s}}, \qquad \, \lim_{r\to\infty} A = -n
,
\label{405}
\end{align}
\endgroup
where the boundary condition of scalar amplitude at infinity is consistent with localized charge distribution, 
$\displaystyle{\lim_{r\rightarrow\infty}\rho=0}$ from \eqref{215}, a necessary condition of finite charge per unit length along $z$ axis \eqref{315}.
Similarly, the boundary conditions of scalar potential and equivalently the neutral scalar field are also assigned,
\begin{align}
\lim_{r\to 0} \frac{d \Phi}{dr} =\lim_{r\to 0} \frac{dN}{dr} = 0 
, \qquad\quad\: 
\lim_{r\to\infty} \frac{d\Phi}{dr} =\lim_{r\to\infty} \frac{dN}{dr} = 0
.
\label{461}
\end{align}

 Cylindrically symmetric vortex solutions have $r$-dependent magnetic field expressed by nontrivial profile of $A(r)$,
\begin{align}
B (r) = \frac{\hbar}{q} \frac{1}{r}\frac{dA(r)}{dr}
,
\end{align}
and their magnetic fluxes \eqref{402} are calculated again by the boundary values of $A(r)$ in \eqref{415}--\eqref{405},
\begin{align}
\Phi_{B} = \frac{2\pi\hbar}{q}[A(\infty)-A(0)] = -2\pi\Phi_{{\rm L}}n
.\label{401}
\end{align}
Since the obtained quantized magnetic flux in \eqref{402}  is always valid for the configurations of arbitrary shape without cylindrical symmetry \eqref{401} and independent of the quartic self-interaction coupling $\lambda$, each vortex of unit vorticity carries the unit magnetic flux $-2\pi \Phi_{\rm L} = 2.07 \times 10^{-15} \, \text{Wb}$ for the Cooper pair.

Any static extended object with nontrivial amplitude of the Schr{\"o}dinger type complex scalar matter means 
$|\Psi|(\boldsymbol{x}) \neq \sqrt{n_{\rm s}}~\,\text{for some planar positions }\boldsymbol{x}$, and this
results in a nonzero localized charge distribution \eqref{215}. Then the Gauss' law \eqref{209} dictates nonzero electric field, $\boldsymbol{E} \neq 0$. In case of  the topological vortex solution of our interest, its complex scalar amplitude connects monotonically the boundary values in 
\eqref{415}--\eqref{405} as we shall see in the subsection 4.3. The charge density of it \eqref{215} is nonnegative everywhere and hence the charge per unit length along $z$ axis \eqref{315} is always nonzero positive, that means the topological vortex of our interest  is hard to be left as an 
electrically neutral object.
When a vortex of our interest carries the net finite electric charge per unit length along $z$ axis $\bar{Q}_{\text{U}(1)}$ accumulated in a local region, it leads to nonzero radial component of the electric field at spatial asymptote of large distance $r$ of $xy$ plane,
\begin{align}
E_{r} = -\frac{d\Phi}{dr} 
\approx \frac{\bar{Q}_{\text{U}(1)}}{2 \pi \epsilon_{0}}
\frac{1}{r}
.\label{443}
\end{align}

We explore the vortex solutions by dividing into the following three cases in the remaining part of the current section.
First, we consider the case that phonon is decoupled from the vortex by turning off the cubic Yukawa type coupling $g=0$.
Second, we consider the case that phonon is coupled to the vortex by turning on the cubic Yukawa type coupling, $ g < g_{c} $ and $ g_{c} < g < 0 $, except the critical coupling.
Lastly, we investigate the case of the critical cubic Yukawa type coupling $ g = g_{c} $.

\subsection{$g=0$}

When the cubic Yukawa type coupling $g$ is turned off, $g=0$, the neutral scalar field of acoustic phonon $N$ decouples  from the effective action \eqref{201} and we
finally have\begingroup
\allowdisplaybreaks
\begin{align}
S_{0} =
& 
\int dt \int d^{2} \boldsymbol{x} \int dz \,
\Big[
- \frac{ \epsilon_{0} c^{2} }{4}
F_{\mu\nu} F^{\mu\nu}
+ q n_{\rm s} \Phi+ \frac{i \hbar }{2} (\bar{\Psi} \mathcal{D}_{t} \Psi 
- \overline{\mathcal{D}_{t} \Psi} \Psi ) 
- \frac{\hbar^{2}}{2m} \overline{ \mathcal{D}_{i}\Psi}\mathcal{D}_{i}\Psi
\nonumber\\
&~~~~~~~~~~~~~~~~~~~~~~~~~
-\lambda (|\Psi|^{2} - v^{2})^{2} 
\Big]
,
\label{465}
\end{align}
\endgroup
where the decoupled neutral scalar field sets to be zero $N=0$ without losing generality.
Then the energy per unit length along $z$ axis \eqref{361} for static field configurations symmetric under the translation along $z$ axis is reduced to
\begingroup
\allowdisplaybreaks
\begin{align}
\bar{E}_{0}
& = \int d^{2} \boldsymbol{x} \,
\bigg\{
\frac{\epsilon_{0} }{2} (\boldsymbol{E}^{2}+c^{2}B^{2}
)+ \frac{\hbar^{2}}{2m}(\boldsymbol{\nabla} |\Psi|)^{2}
+ \frac{q^{2}}{2m} |\Psi|^{2}\Big( A^{i} - \frac{\hbar}{q} \partial_{i} \Omega \Big)^{2}
+ \lambda (|\Psi|^{2} - v^{2})^{2}
\nonumber\\
& \hspace{4.8em}
+\boldsymbol{\nabla} \cdot ( \epsilon_{0} \Phi \boldsymbol{E} )
- \epsilon_{0} \Phi \Big[ \boldsymbol{\nabla} \cdot \boldsymbol{E} 
- 
\frac{q}{\epsilon_{0}} (|\Psi|^{2} - n_{\rm s}) \Big]
\bigg\}
.\label{436}
\end{align}
\endgroup

Suppose that there exists a static object of zero charge and additionally no electric field is accompanied, $\boldsymbol{E}={\bf 0}$. Then the corresponding scalar potential can always be chosen by zero, $\Phi=0$. Substitution of these into \eqref{436} gives
\begin{align}
\bar{E}_{0}
& = \int d^{2} \boldsymbol{x} \,
\bigg[
\frac{\epsilon_{0}c^{2}}{2}B^{2}
+ \frac{\hbar^{2}}{2m}(\boldsymbol{\nabla} |\Psi|)^{2}
+ \frac{q^{2}}{2m} |\Psi|^{2}\Big( A^{i} - \frac{\hbar}{q} \partial_{i} \Omega \Big)^{2}
+ \lambda (|\Psi|^{2} - v^{2})^{2}
\bigg]
,
\end{align}
whose formula is exactly the same as the Ginzburg-Landau free energy for the electrically neutral static Abrikosov-Nielsen-Olesen vortices. As a result of this discussion, one may regard the action \eqref{465} as the mother effective field theory of conventional superconductivity including a Schr\"{o}dinger type complex scalar field of Cooper pair.
As already discussed in the section 2, any classical configuration is restricted by the Gauss' law \eqref{209} with the charge density \eqref{215}. The constant symmetry-broken vacuum in \eqref{221} is automatically the symmetry-broken vacuum of type I superconducting state with minimum zero energy of this action, in which the Gauss' law is trivially satisfied. 

If the vortices of $n\ne0$ are taken into account, the nontrivial vortex solution obeying the boundary conditions \eqref{415}--\eqref{405} leads to a nonzero electric charge per unit length along $z$ axis $\bar{Q}_{{\rm U(1)}}\ne0$, the Coulombic electric field \eqref{443} is developed, and hence the scalar potential
develops a logarithmic divergence for sufficiently large $r$,
\begin{align}
\Phi \approx - \frac{ \bar{Q}_{{\rm U}(1)} }{ 2\pi \epsilon_{0} } 
\ln \frac{r}{\lambda_{\rm L}}
.
\label{462}
\end{align}  
Substitution of it into the energy per unit length along $z$ axis \eqref{436} results in negative logarithmic divergent energy in $xy$ plane from the divergence term,
\begin{align}
\bar{E}_{0}\approx \lim_{R\rightarrow\infty}-\frac{\bar{Q}_{{\rm U(1)}}^{2}}{4\pi\epsilon_{0}}\ln 
\frac{R}{\lambda_{{\rm L}}}.
\end{align}
Even if this logarithmic divergence is considered to be allowable by the energetics of probable generation in quantum regime, the key obstacle is nonexistence of such nonsingular vortex solution.
If we substitute the scalar potential with logarithmic divergence \eqref{462} into the scalar amplitude equation
\begin{align}
\frac{d^{2} |\Psi|}{dr^{2}} + \frac{1}{r} \frac{d|\Psi|}{dr} 
 = \frac{(A+n)^{2}}{r^{2}} |\Psi| 
+ \frac{1}{2\xi^{2}} \Big(\frac{|\Psi|^{2}}{n_{\rm s}} - 1\Big) |\Psi| + \frac{q}{4\lambda n_{\rm s}\xi^{2} }  
\Phi |\Psi|
,
\label{463}
\end{align}
and try the expansion of scalar amplitude at large $r$, the last term in the right-hand side of the equation \eqref{463} includes logarithmic divergence and such divergence leads to inconsistency contrary to the nonzero finite boundary value of the scalar amplitude
$ \displaystyle
\lim_{r\to\infty} |\Psi|(r)
= \sqrt{n_{\rm s}}
$
in \eqref{405}. Hence no nonsingular vortex solution is supported in the field theory of the action \eqref{465} and furthermore its BPS limit is forbidden. It means in conventional superconductivity described by this action \eqref{465} that there is no type I$\!$I superconductors including the borderline between type I and I$\!$I superconductors. Therefore, the field theory of the action \eqref{465} without neutral scalar field is irrelevant to the Ginzburg-Landau theory of the Ginzburg-Landau free energy which first predicted the regular vortex solutions for type I$\!$I superconductors and their BPS limit for the borderline of type I and I$\!$I superconductors.

Before moving to the next case, we recapitulate the Abelian Higgs model. It supports the same symmetry-broken vacuum of type I superconductivity and the same Abrikosov-Nielsen-Olesen vortices as regular static solutions of finite energy \cite{Abrikosov:1956sx, Nielsen:1973cs}.
The static vortices in the Abelian Higgs model are electrically neutral. In the Gauss' law, the time-independent charge density is proportional to the scalar potential, or equivalently the time component of U(1) gauge field $A^{0}$,
\begin{align}
\rho \propto A^{0} \big[|\Psi|(\boldsymbol{x})^{2} - v^{2}\big]= \frac{\Phi}{c} \big[|\Psi|(\boldsymbol{x})^{2} - v^{2}\big]
.
\label{468}
\end{align}
Even any static object of arbitrary distribution of nonzero matter density, $ |\Psi|(\boldsymbol{x})^{2} - v^{2} \neq 0 $, can have zero charge density everywhere $\rho(\boldsymbol{x}) = 0$ that can always be implemented by a consistent choice of the Weyl gauge fixing condition $A^{0} = 0$ ($\Phi=0$). For the critical coupling of quartic self-interaction, the BPS bound is saturated\cite{Bogomolny:1975de} and there exist both type I and I$\!$I superconductors. Now there remains only one question that bothers us to accept it as the field theory of conventional superconductivity: Is the relativistic complex scalar field with the light speed $c$ as the characteristic propagation speed appropriate as the field of a Cooper pair, a composite field of two electrons, in a condensed matter sample in superconducting phase? A simple way out of this relativistic regime is to replace the light speed to a nonrelativistic characteristic speed, $c\rightarrow v_{{\rm p}}$, in the Abelian Higgs model \cite{Kim:2024gfn, Jeon:2025snd}. Though every aforementioned properties of the relativistic Abelian Higgs model are maintained,
we are left with a homework in experiments to measure this characteristic propagation speed $v_{{\rm p}}$ of the Cooper pair. 

In this subsection 4.1, we showed that the field theory of the action \eqref{465} does not support the static regular vortex solutions of finite energy. Let us turn on the cubic Yukawa type coupling $g\ne0$ between neutral and complex scalar field, and continue investigation of the vortex solutions.

\subsection{$g<g_{c}$ or $g_{c}<g<0$}

When the cubic Yukawa type coupling $g$ is turned on $g\ne0$ but not critical value $g \neq g_{\rm c}$ in \eqref{327}, the neutral scalar field of acoustic phonon $N$ is coupled to the Cooper pair as in  the $(1+2)$-dimensional action 
\eqref{301}.
The equation of vector potential \eqref{418}  and the formula of scalar potential and neutral scalar field \eqref{471} are unchanged but the equation of scalar amplitude \eqref{470} is written without the neutral scalar field by use of
\begin{align}
\frac{d^{2} |\Psi|}{dr^{2}} + \frac{1}{r} \frac{d|\Psi|}{dr} 
= \frac{(A+n)^{2}}{r^{2}} |\Psi| 
+ \frac{1}{2\xi^{2}} \Big(\frac{|\Psi|^{2}}{n_{\rm s}} - 1\Big) |\Psi| - \frac{\epsilon_{0}}{4q\lambda n_{\rm s} \xi^{2}} \Big( g^{2}- \frac{q^2}{\epsilon_{0}} \Big) \Phi |\Psi| 
,
\label{407}
\end{align}
where the neutral scalar field $N(r)$ is eliminated by use of the relation \eqref{310} in the last term of the equation \eqref{407} with zero residual gauge degrees of freedom $f=0$. Since the coefficient of the last term of \eqref{407} is not zero, $g^{2}-q^{2}/\epsilon_{0}\ne0$, the scalar equation of the current case \eqref{407} and the previous one \eqref{463} take mathematically the same form. Therefore, we reach the nonexistence of nonsingular static vortex solutions of finite energy according to the same mathematical logic of the $g=0$ case. 

Nonexistence of nonsingular vortex solution is established for both zero cubic Yukawa type coupling $g=0$ in the previous subsection 4.1 and nonzero but noncritical cubic Yukawa type coupling $g<g_{c}$ or $g_{c}<g<0$ in the current subsection 4.2.
Note that there still remains a possible phase materialized through the formation of vortex-antivortex pairs with zero net vorticity.
Similar to the phase by global $\mathrm{U}(1)$ vortex-antivortex pairs in the Kosterlitz-Thouless theory \cite{Kosterlitz:1973xp}, this phase by the gauged $\mathrm{U}(1)$ vortex-antivortex pairs of zero net electric charge is safe from infinite energy and singularity of the solutions.
The only remaining case for vortex or antivortex solutions of net nonzero vorticity is that of critical cubic Yukawa type coupling $g=g_{c}$ discussed in the next subsection.

\subsection{$g=g_{c}$}

When cubic Yukawa type coupling takes the critical value $g=g_{{\rm c}}$ \eqref{327}, the terms in the last line of the energy per unit length along $z$ axis \eqref{460} and \eqref{417} become exactly zero for any static solution together and all the terms in the second line do with the help of the relation \eqref{310} with $f=0$.
Hence only the four terms in the first line are left,
\begingroup
\allowdisplaybreaks
\begin{align}
\bar{E}& = \int d^{2} \boldsymbol{x} \,
\bigg[
\frac{\epsilon_{0} c^{2}}{2} B^{2}
+ \frac{\hbar^{2}}{2m} (\boldsymbol{\nabla} |\Psi|)^{2}
+ \frac{q^{2}}{2m} |\Psi|^{2}\Big( A^{i} - \frac{\hbar}{q} \partial_{i} \Omega \Big)^{2}
+ \lambda (|\Psi|^{2} - v^{2})^{2}
\bigg]
.\label{439}
\end{align} 
\endgroup
For cylindrically symmetric configurations, it is
\begingroup
\allowdisplaybreaks
\begin{align}
\bar{E} &
= 2\pi \int_{0}^{\infty} dr \, r \,
\bigg[
\frac{\epsilon_{0} c^{2}\hbar^{2}}{2q^{2}}\frac{1}{r^{2}}
\Big(  \frac{dA}{dr} \Big)^{2}
+ \frac{\hbar^{2}}{2m} \Big( \frac{d|\Psi|}{dr} \Big)^{2}
+ \frac{\hbar^{2}}{2m} \frac{(A+n)^{2}}{r^{2}}  |\Psi|^{2}
+ \lambda (|\Psi|^{2} - v^{2})^{2}
\bigg]
.
\end{align} 
\endgroup
Notice that the terms in \eqref{439} coincide exactly with those of the Ginzburg-Landau free energy \cite{Abrikosov:1956sx} and the energy of the relativistic and nonrelativistic Abelian Higgs model \cite{Nielsen:1973cs, Kim:2024gfn}.
Accordingly the unchanged gauge field equation \eqref{418} and the complex scalar equation decoupled from the neutral scalar field are equal to those of the Ginzburg-Landau theory and the Abelian Higgs model,
\begin{align}
\frac{d^{2} |\Psi|}{dr^{2}} + \frac{1}{r} \frac{d|\Psi|}{dr} 
= \frac{(A+n)^{2}}{r^{2}} |\Psi| 
+ \frac{1}{2\xi^{2}} \Big(\frac{|\Psi|^{2}}{n_{\rm s}} - 1\Big) |\Psi|
.
\end{align}
Thus every vortex profiles of the scalar amplitude and the gauge field always take the same functional form as those  of the Abrikosov-Nielsen-Olesen vortices and the values of corresponding energy of static $n$ vortices are finite and equal irrespective of the quartic self-interaction coupling $\lambda$. It implies that no distinction is observed between our charged vortices and the neutral Abrikosov-Nielsen-Olesen vortices as long as the magnetic field or the Cooper pairs are measured in experiments. 
However, the charged vortices of our effective field theory are completely different from the neutral Abrikosov-Nielsen-Olesen vortices whose electric charges are originated from their Gauss' law \eqref{209} with different charge densities. To be specific, nonzero and nontrivial \eqref{215} in our effective field theory and zero \eqref{468} with the help of Weyl gauge $\Phi=0$ in the Ginzburg-Landau theory and Abelian Higgs model, respectively. Though almost all the profiles of $|\Psi|$ and $A$ are given in the textbooks
\cite{Jacobs:1978ch,Manton:2004tk,Weinberg:2012pjx}, we reproduce them in order to study the scalar potential, the electric field, and the angular momentum density with $\lambda$-dependence of charge.

The boundary conditions \eqref{415}, \eqref{405}, and \eqref{461} are assigned without change.
Power series expansion of the scalar amplitude and the gauge field near the origin gives,
\begingroup
\allowdisplaybreaks
\begin{align}
|\Psi|(r) & \approx
\sqrt{n_{\rm s}} \Psi_{n} \Big( \frac{r}{\xi} \Big)^{|n|}
\bigg\{
1 +\frac{1}{8(|n|+1)}
\Big( \frac{4n A_{n}}{\kappa^{2}} - 1 \Big)
\Big( \frac{r}{\xi} \Big)^{2} \nonumber
\\ 
& 
+ \frac{1}{8(|n|+2)}
\bigg[
\frac{A_{n}^{2}}{\kappa^4}
+ \frac{1}{4(|n|+1)}
\Big( \frac{4n A_{n}}{\kappa^2} + 1 \Big)
+ \frac{1 + 2\kappa^{2}}{4\kappa^{2}} \Psi_{n}^{2} \delta_{|n|1}
\bigg]
\Big( \frac{r}{\xi} \Big)^{4} + \cdots
\bigg\}
, 
\label{408}\\
A(r) & \approx
A_{n} \Big( \frac{r}{\lambda_{\rm L}} \Big)^{2}
+ \frac{n \kappa^{2|n|} |\Psi_{n}|^{2}}{4|n|(|n|+1)} \Big( \frac{r}{\lambda_{\rm L}} \Big)^{2|n| + 2}
+ \cdots,
\label{409}
\end{align}
\endgroup
where $\Psi_{n}$ and $A_{n}$ are two dimensionless constants and have different values for different vorticity $n$.
Substitution of the small $r$ expansion \eqref{408} into \eqref{473} leads to
\begin{align}
N(r) &= - \sqrt{\epsilon_0} \Phi (r)
\nonumber\\
&
\approx
\frac{2 \sqrt{\epsilon_0} \lambda n_{\rm s}}{q}
\bigg[ N_{n} 
- \frac{ q^{2} \xi^{2} }{8 \epsilon_{0} \lambda} 
\Big( \frac{r}{\xi} \Big)^{2}
+ \frac{q^{2} \xi^{2}\Psi_{n}^{2}}{8 \epsilon_{0} \lambda(|n|+1)^{2}}
\Big( \frac{r}{\xi} \Big)^{2|n|+2} + \cdots
\bigg]
, 
\label{410}
\end{align}
where an integration constant $N_n$ is a constant zero mode along the flat direction of the scalar potential \eqref{206}, whose value is appropriately determined according to the fixation of the gauge degree of freedom.
Asymptotic solutions of the $n$ vortices superimposed at the origin are
\begingroup
\allowdisplaybreaks
\begin{align}
|\Psi|(r) &
\approx \sqrt{n_{\rm s}} \bigg[ 1 - \Psi_{\infty} K_{0}  \Big(\frac{r}{\xi} \Big) \bigg]
,\label{441}
\\
A(r) &\approx 
- n - A_{\infty} \Big(\frac{r}{\lambda_{\rm L}}\Big) K_{1} \Big(\frac{r}{\lambda_{\rm L}}\Big)
,
\end{align}
\endgroup
where dimensionless constants $\Psi_{\infty}$ and $A_{\infty}$ are related to the constants near the origin, $\Psi_{n}$ and $A_{n}$.
Since the scalar amplitude \eqref{441} approaches exponentially the boundary value \eqref{405}, the integrated electric charge per unit length along $z$ axis $ \bar{Q}_{\mathrm{U}(1)} $ \eqref{315} is finite.
Therefore, the neutral scalar field has logarithmically divergent leading term for sufficiently large $r$
and the corresponding scalar potential determined by the relation \eqref{310} with $f=0$ is also logarithmic at large distance
\begin{align}
\Phi =-\frac{q}{\epsilon_{0}g}N\approx
- \frac{\bar{Q}_{\text{U}(1)}}{2 \pi \epsilon_{0}}
\bigg[
\ln \frac{r}{\lambda_{\rm L}}
- \frac{q^{2} \xi^{2}}{\epsilon_{0} \lambda} \Psi_{\infty} K_{0}  \Big(\frac{r}{\xi} \Big)
\bigg]
,\label{432}
\end{align}
which means a marginally confining potential as usual in $(1+2)$-dimensional electrostatics.

\begin{figure}[H]
\centering
\subfigure[]{\includegraphics[width=0.44\textwidth]{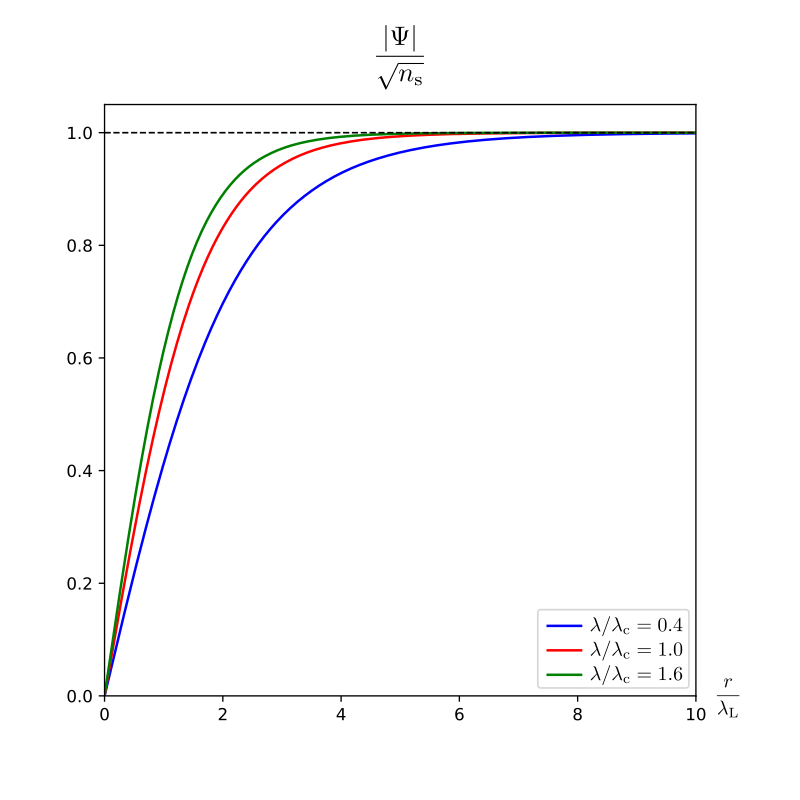}}
\hfill
\subfigure[]{\includegraphics[width=0.44\textwidth]{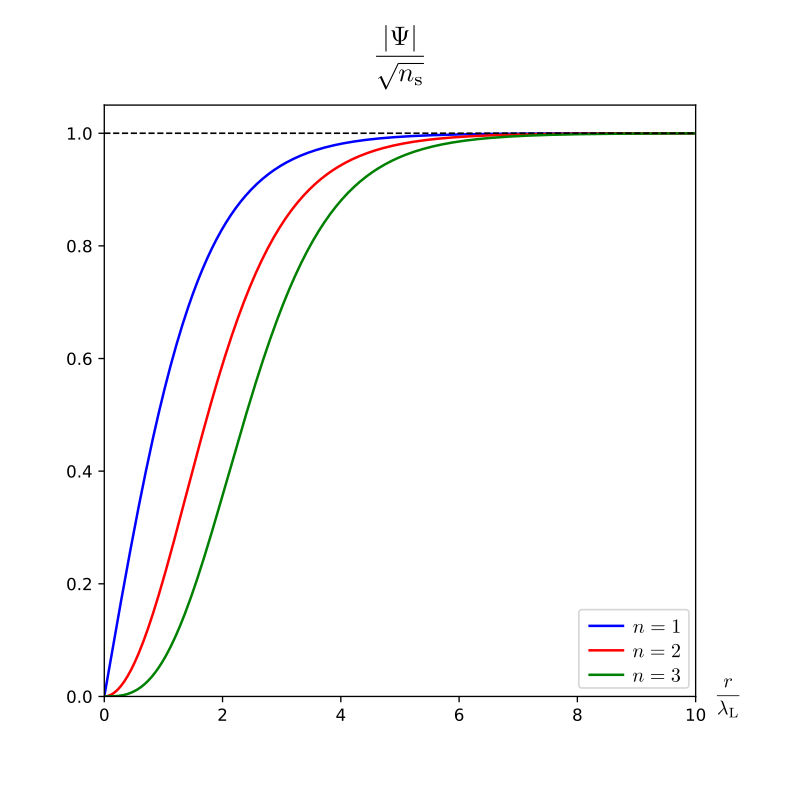}}
\vskip\baselineskip
\subfigure[]{\includegraphics[width=0.44\textwidth]{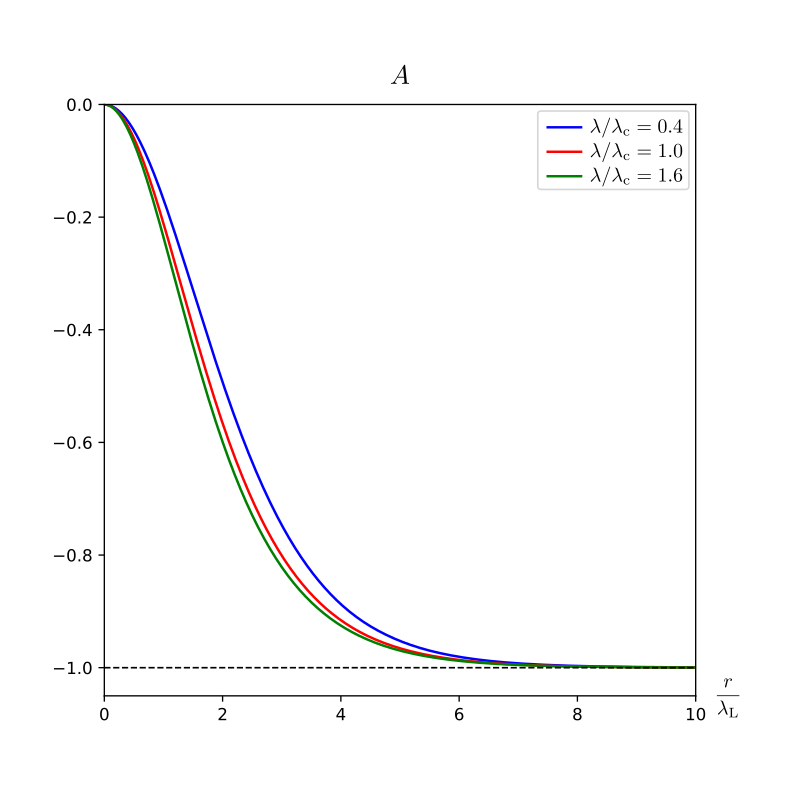}}
\hfill
\subfigure[]{\includegraphics[width=0.44\textwidth]{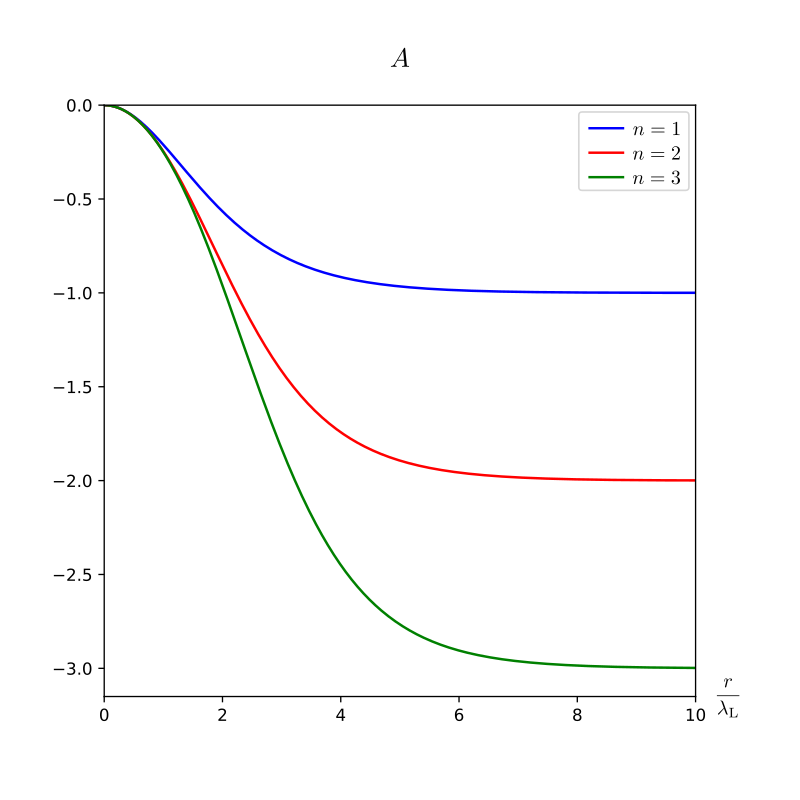}}
\caption{Profiles of the scalar amplitude $|\Psi|$ in the unit of $\sqrt{n_{\rm s}}$ for (a) $ \lambda / \lambda_{\rm c} = 0.4, 1.0, 1.6 $ with $n=1$ and (b) $n=1,2,3$ with $\lambda = \lambda_{\rm c}$, as functions of a dimensionless length variable $r/\lambda_{\rm L}$. Profiles of the angular component of gauge field $A$ for (c) $ \lambda / \lambda_{\rm c} = 0.4, 1.0, 1.6 $ with $n=1$ and (d) $n=1,2,3$ with $\lambda = \lambda_{\rm c}$, as functions of a dimensionless length variable $r/\lambda_{\rm L}$.}
\label{fig:402}
\end{figure}

\begin{figure}[H]
\centering
\subfigure[]{\includegraphics[width=0.48\textwidth]{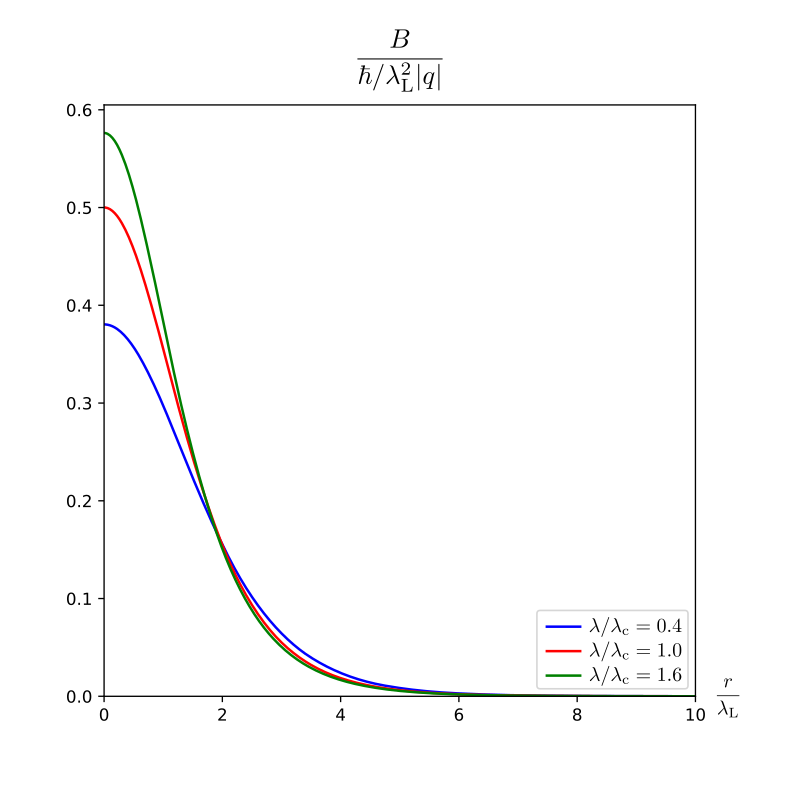}}
\hfill
\subfigure[]{\includegraphics[width=0.48\textwidth]{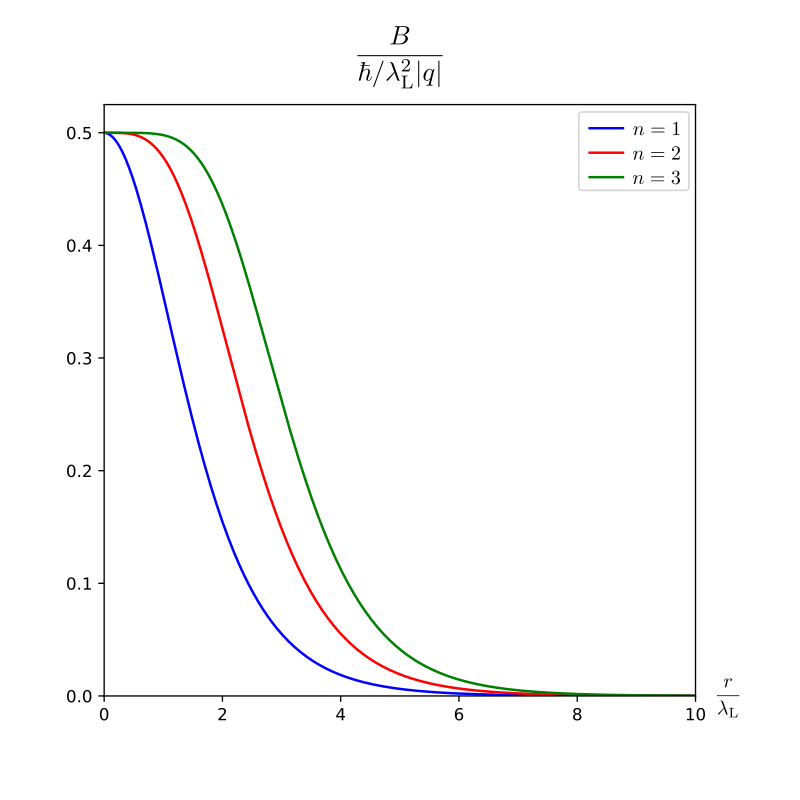}}
\caption{Profiles of the magnetic field $B$ in the unit of $ \hbar / \lambda_{\rm L}^{2} |q| $ for (a) $\lambda / \lambda_{\rm c} = 0.4, 1.0, 1.6$ with $n=1$ and (b) $n=1,2,3$ with $\lambda = \lambda_{\rm c}$, as functions of a dimensionless length variable $ r / \lambda_{\rm L} $.}
\label{fig:403}
\end{figure}

Since no exact vortex solution is obtained even in the BPS limit of critical coupling $\lambda = \lambda_{\rm c}$ with a single scalar equation \eqref{428}, numerical analysis is employed.
Figure \ref{fig:402} shows well-known cylindrically symmetric profiles of the vortex configurations of unit vorticity $n=1$ for various quartic self-interaction coupling $\lambda$ as in figure \ref{fig:402}-(a) and (c) and of vorticities $n=1,2,3$ with fixed coupling $\lambda = \lambda_{\rm c}$ as in figure \ref{fig:402}-(b) and (d).
The scalar amplitude $|\Psi|(r)$ and the angular component of gauge field $A(r)$ connect smoothly the boundary values \eqref{415}--\eqref{405}.
As the quartic self-interaction coupling $\lambda$ increases, radial sizes of the vortex solutions decrease as shown in figure \ref{fig:402}-(a) and (c).
The obtained numerical data are consistent with the correlation length \eqref{303} shorter than the London penetration depth \eqref{302} and attractive nature of the short ranged interaction mediated by the massive Higgs field.
When vorticity increases $n=1,2,3$, radial sizes of the vortex solutions increase as shown in figure \ref{fig:402}--(b), that is consistent with $n$ vortices superimposed at a position, the origin.
\noindent
Nontrivial short ranged profiles of the angular component of vector potential $A(r)$ mean nontrivial magnetic fields along $z$ axis $B(r)$ whose directions are upward for the vortices of positive vorticities $n\ge1$ and downward for the vortices of negative vorticities $n\le-1$ as shown in figure \ref{fig:403}.
As the quartic self-interaction coupling $\lambda$ increases, magnetic field becomes more concentrated at the center with higher peak at $r=0$ as shown in figure \ref{fig:403}-(a).
When vorticity $n$ grows, magnetic fields spread radially as shown in figure \ref{fig:403}-(b).
Though the shapes of magnetic fields vary continuously on the quartic self-interaction coupling $\lambda$, the magnetic flux $\Phi_B$ \eqref{401} is left to be proportional to discrete vorticity $n$ but independent of the quartic self-interaction coupling.

Suppose that a constant external magnetic field $\boldsymbol{B}^{\rm ext} = (0,0,B^{\rm ext})$ is applied to the flat $xy$ plane of a superconducting sample with translation symmetry along $z$ axis.
Since any static vortex solution exists in the critical cubic Yukawa type coupling $g=g_{\rm c}$, the energy per unit length along $z$ axis reduces with the help of symmetric gauge to
\begingroup
\allowdisplaybreaks
\begin{align}
\bar{E}
&= \int d^{2} \boldsymbol{x} \,
\bigg[ \frac{\epsilon_{0} c^{2}}{2} (B - B^{\rm ext})^{2}
+ \frac{\hbar^{2}}{2m} (\partial_{i}|\Psi|)^{2} + \frac{q^{2} |\Psi|^{2}}{2m} \Big( A^{i} - \partial_{i} \Omega + \frac{1}{2} \epsilon^{ij} x_{j} B^{\rm ext} \Big)^{2} \nonumber\\
&~~~~~~~~~~~~~~ + \lambda (|\Psi|^{2} - v^{2} )^{2}
\bigg]
.\label{412}
\end{align}
\endgroup
If the $\text{U}(1)$ gauge field cancels perfectly the external magnetic field,
\begin{align}
\Phi \rightarrow \Phi, \quad A^{i} \rightarrow A^{i} - \frac{1}{2} \epsilon^{ij} x_{j} B^{\rm ext}
,\label{411}
\end{align}
and the scalar and vector potentials $\Phi$ and $A^{i}$ in \eqref{411} describe a static vortex solution without the external magnetic field, it also satisfies the Euler-Lagrange equations in the presence of the constant external magnetic field and then the energy per unit length along $z$ axis \eqref{412} becomes equal to that without the external magnetic field.
Though the external magnetic field is canceled, there remains nonzero short ranged magnetic field at every vortex site.
It means the penetration of magnetic field through the superconducting sample and this Meissner effect in type I$\!$I superconductors including vortices explains imperfect diamagnetism and is understood as an energy minimization procedure for the given topological sector of winding number $n$.

\begin{figure}[H]
\centering
\subfigure[]{\includegraphics[width=0.48\textwidth]{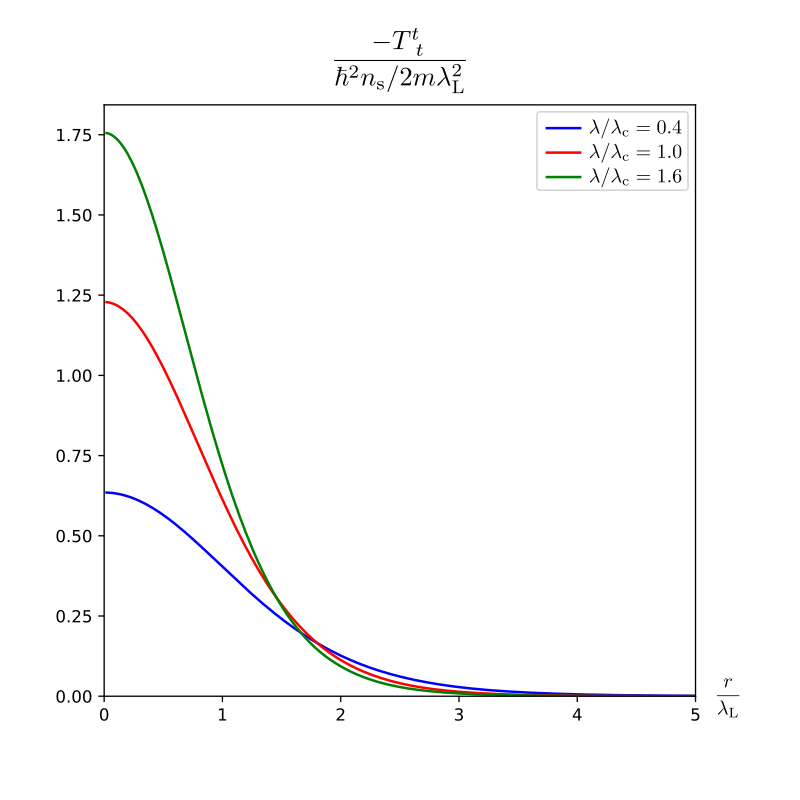}}
\hfill
\subfigure[]{\includegraphics[width=0.48\textwidth]{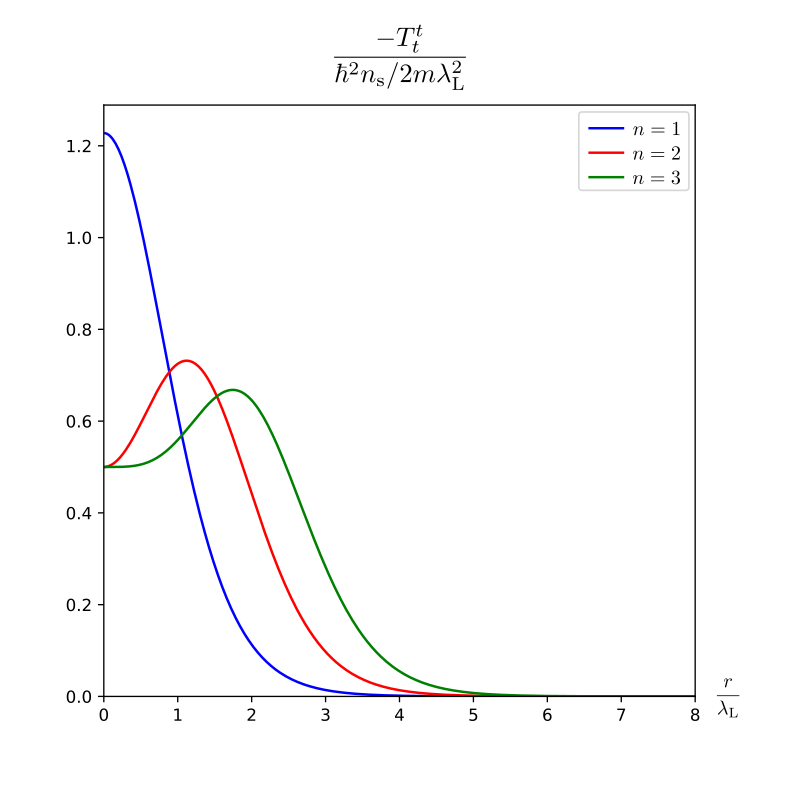}}
\caption{Profiles of the energy density $ -T \indices{^t_t} $ in the unit of its characteristic scale $ \hbar^{2} n_{\rm s} / 2 m \lambda_{\rm L}^{2} $ for (a) $ \lambda / \lambda_{\rm c} = 0.4, 1.0, 1.6 $ with $n=1$ and (b) $n=1,2,3$ with $\lambda = \lambda_{\rm c}$, as functions of a dimensionless length variable \(r / \lambda_{\rm L}\).}
\label{fig:404}
\end{figure}

Similar to magnetic field $B$, profiles of energy density $ -T\indices{^t_t} $ \eqref{417} of $n=1$ vortex are distributed within their radial sizes characterized by two length scales, the London penetration depth $\lambda_{\rm L}$ \eqref{302} and the correlation length $\xi$ \eqref{303}.
As the quartic self-interaction coupling $\lambda$ increases, energy density is more concentrated near the origin with an increased peak at the origin as shown in figure \ref{fig:404}-(a).
As vorticity increases, energy densities of $n=1, 2, 3$ vortices change their shapes from a distribution with maximum peak at the origin for $n=1$ to ring-shapes for $n\ge2$ and spread radially outward consistent with increasing energy proportional to the vorticity \eqref{421} as shown in figure \ref{fig:404}-(b).
\noindent
In the section~2, the limit for the borderline of type I and I$\!$I superconductors is introduced by the equality \eqref{314} between the correlation length \eqref{303} and the London penetration depth \eqref{302}, which equivalently leads to the critical value of quartic self-interaction coupling $\lambda_{\rm c}$ \eqref{312} \cite{Abrikosov:1956sx}.
We nonperturbatively confirmed this limit as the BPS limit \eqref{421} in which noninteracting multi-vortices with arbitrary interdistances are supported, while they are charged.
Subsequently, by using perturbative analysis, the scattering amplitudes of two Cooper pairs calculated at $1$-loop level are taken into account, and it is understood in terms of the resultant zero net interaction between the attraction mediated by the massive Higgs and the repulsion mediated by the massive mode of $\text{U}(1)$ gauge field in the static limit \cite{NRAH}.

\begin{figure}[H]
\centering
\subfigure[]{\includegraphics[width=0.48\textwidth]{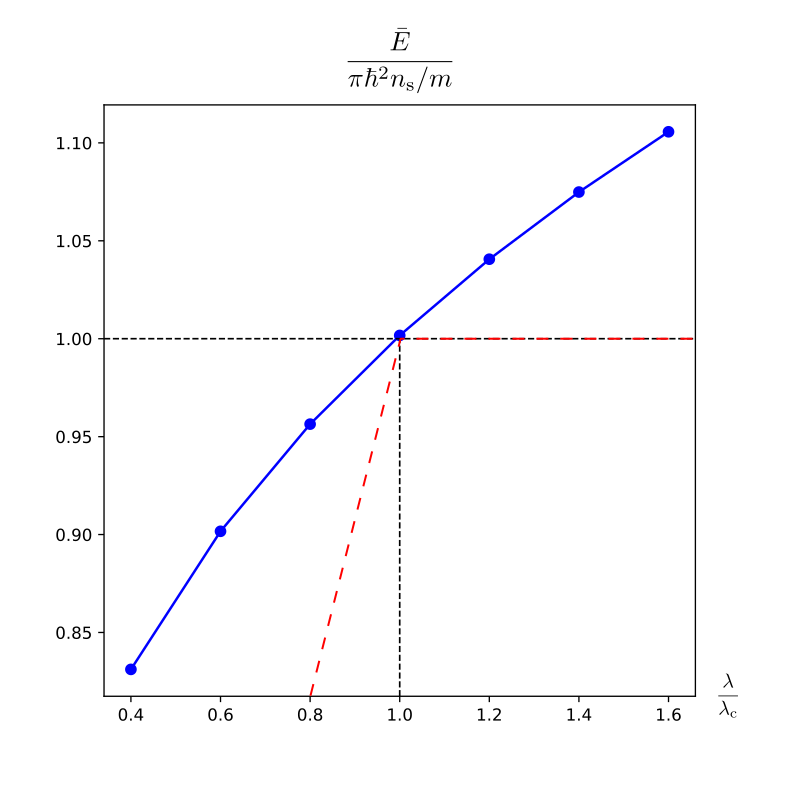}}
\hfill
\subfigure[]{\includegraphics[width=0.48\textwidth]{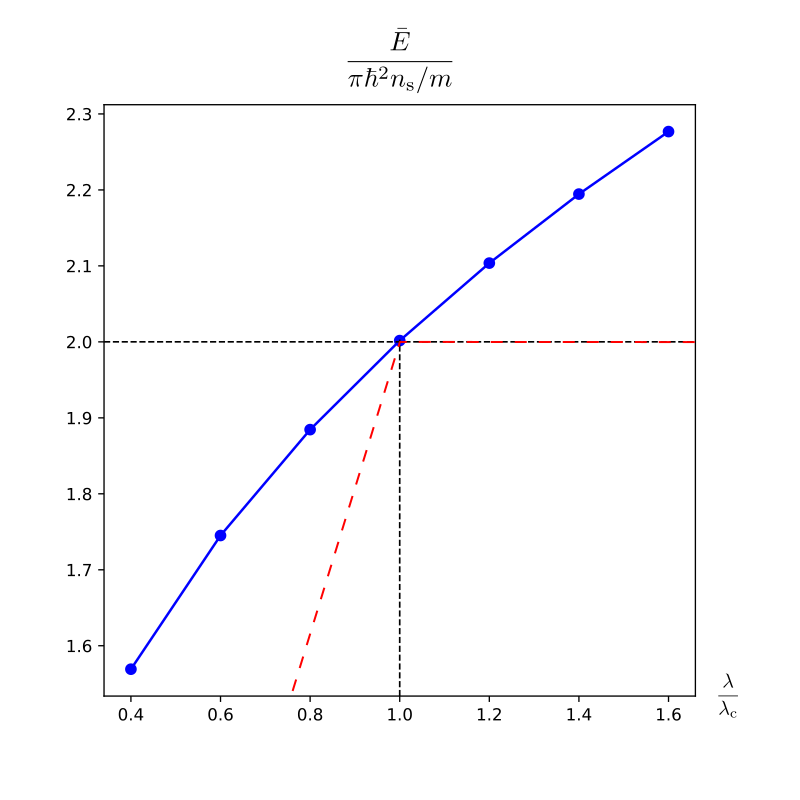}}
\caption{The energy per unit length along \(z\)-axis integrated inside a cylindrical region \( r < 20 \lambda_{\rm L} \), of (a) an \(n=1\) vortex and (b) $n=2$ for various values of the quartic scalar coupling \(\lambda\), in the unit of \( \pi \hbar^{2} n_{\rm s} / m \).}
\label{fig:406}
\end{figure}

\begin{figure}[H]
\centering
\subfigure{\includegraphics[width=0.5\textwidth]{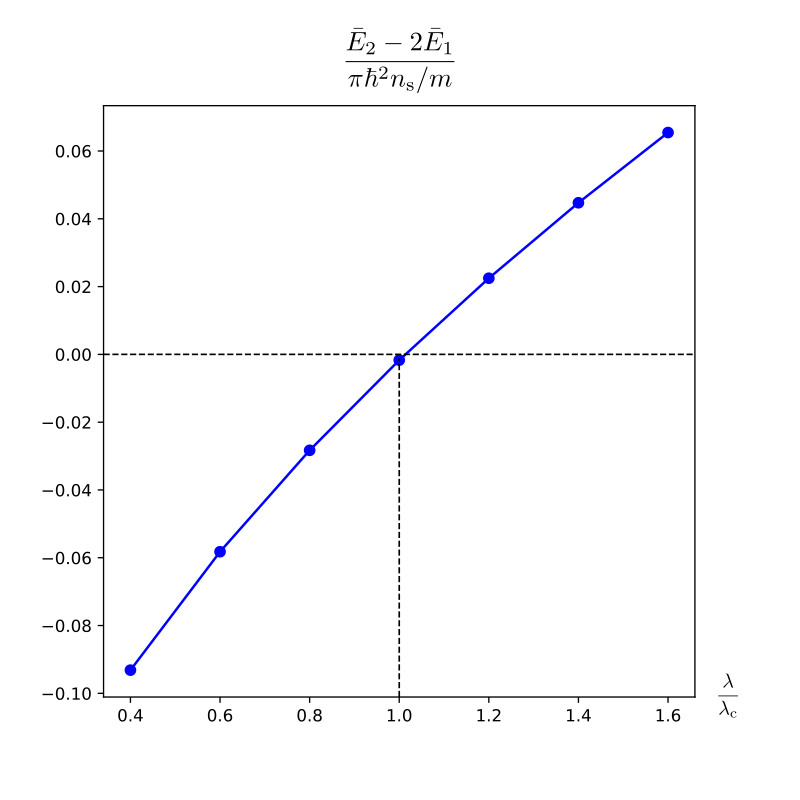}}
\caption{2-body interaction energy per unit length along \(z\)-axis integrated inside a cylindrical region \( r < 20 \lambda_{\rm L} \) for various values of the quartic self-interaction coupling \(\lambda\), in the unit of \( \pi \hbar^{2} n_{\rm s} / m \).}
\label{fig:407}
\end{figure}

It is the turn to discuss type I or I$\!$I superconductivity for the regime of weak or strong quartic self-interaction coupling, in addition to the borderline of critical coupling $\lambda_{{\rm c}}$.
When the quartic self-interaction coupling $\lambda$ is away from the critical coupling, analytic method is applied in a limited manner however some inequalities for energy per unit length along $z$ axis are derived \cite{Jacobs:1978ch, Jeon:2025snd}, i.e., the lower bounds of the energy are achieved including the effects of both neutral scalar field $N$ and background charge density $n_{{\rm s}}$. In the weak coupling regime of $ 0 \le \lambda < \lambda_{{{\rm c}}} $, an inequality holds,
\begin{align}
E & \ge \frac{\lambda}{\lambda_{\rm c}} \frac{\hbar n_{\rm s} }{2m} |q\Phi_{B}| - ( g - g_{\rm c} ) \int d^{2} \boldsymbol{x}\, N(n_{\rm s} - |\Psi|^{2})
, \label{403}
\end{align}
and, in the strong coupling regime of $\lambda_{{{\rm c}}} < \lambda$, another inequality holds,
\begin{align}
E & \ge \frac{\hbar n_{\rm s}}{2m} |q\Phi_{B}| - ( g - g_{\rm c} ) \int d^{2} \boldsymbol{x}\, N(n_{\rm s} - |\Psi|^{2})
,\label{404}
\end{align}
where the common last term vanishes for charged vortex solutions supported only at the critical cubic Yukawa type coupling $g=g_{\rm c}$ \eqref{327}.
The derived inequalities \eqref{403}--\eqref{404} are valid for nonperturbative solitonic spectra and then will be tested for the topological vortices in nonBPS regime including their mutual interactions as shown by the red-colored dashed line in figure \ref{fig:406}.
The energy per unit length along \(z\)-axis of $n=1$ and $2$ vortices integrated inside a cylindrical region \( r < 20\lambda_{\rm L} \) are plotted for various quartic self-interaction coupling $\lambda$ as shown in figure \ref{fig:406}-(a)--(b).
As the quartic self-interaction coupling $\lambda$ becomes strong, vortex energy per unit length along $z$ axis increases monotonically.
In the BPS limit of the critical coupling $\lambda=\lambda_{\rm c}$, values of the energy per unit length along $z$ axis for $n=1$ and $2$ vortices are exactly proportional to vorticity, $ n \times \pi \hbar^{2} n_{\rm s} / m $.
\noindent
The net energy per unit length along $z$ of $n$ superimposed vortices is exactly equal to $n$ times that of a single vortex only when $\lambda = \lambda_{\rm c}$.
Hence the interaction energy of 2-body interaction \( \bar{E}_{2}- 2\bar{E}_{1} \) is evaluated as shown in figure \ref{fig:407}.
Zero interaction energy between two vortices in the BPS limit of critical quartic self-interaction coupling $\lambda = \lambda_{\rm c}$ as already calculated exactly by the BPS bound \eqref{421}.
Positive interaction energy in strong coupling regime $\lambda > \lambda_{\rm c}$ exhibits a repulsion between two superimposed vortices with zero interdistance, and negative interaction energy in weak coupling regime $0<\lambda < \lambda_{\rm c}$ does an attraction.
Even when they are superimposed, their interaction is weak, within $10\%$, $|\bar{E}_{2} - 2 \bar{E}_{1}| / (\pi \hbar^{2} n_{\rm s} / m) < 0.1$, in the neighborhood of critical coupling $|\lambda/\lambda_{\rm c} - 1| < 1$.

Again we emphasize that the discussions up to here are exactly the same as those of the Abrikosov-Nielsen-Olesen vortices in the Ginzburg-Landau theory \cite{Abrikosov:1956sx} and the relativistic and nonrelativistic Abelian Higgs model \cite{Nielsen:1973cs, Kim:2024gfn, Jeon:2025snd}.
The prominent distinction between the Abrikosov-Nielsen-Olesen vortex and our vortex appears in electric charge.
In the context of the Abelian Higgs model or our effective field theory of the action \eqref{201}, a mere difference of the static charge densities with including the overall scalar potential \eqref{468} or without it \eqref{215} produces the neutral spinless Abrikosov-Nielsen-Olesen vortex or our charged vortex, respectively.

\begin{figure}[H]
\centering
\subfigure[]{\includegraphics[width=0.48\textwidth]{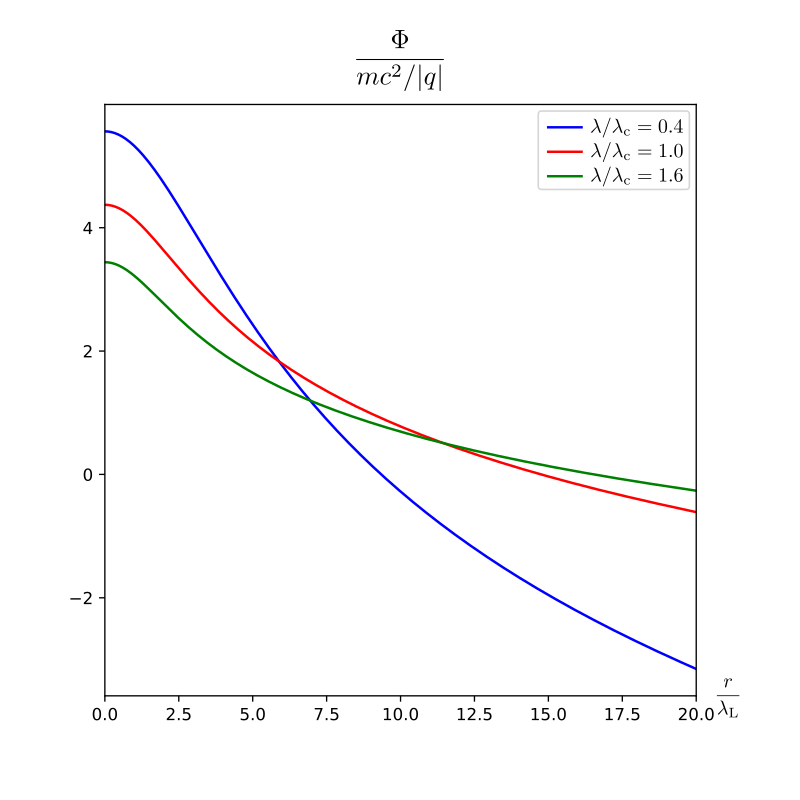}}
\hfill
\subfigure[]{\includegraphics[width=0.48\textwidth]{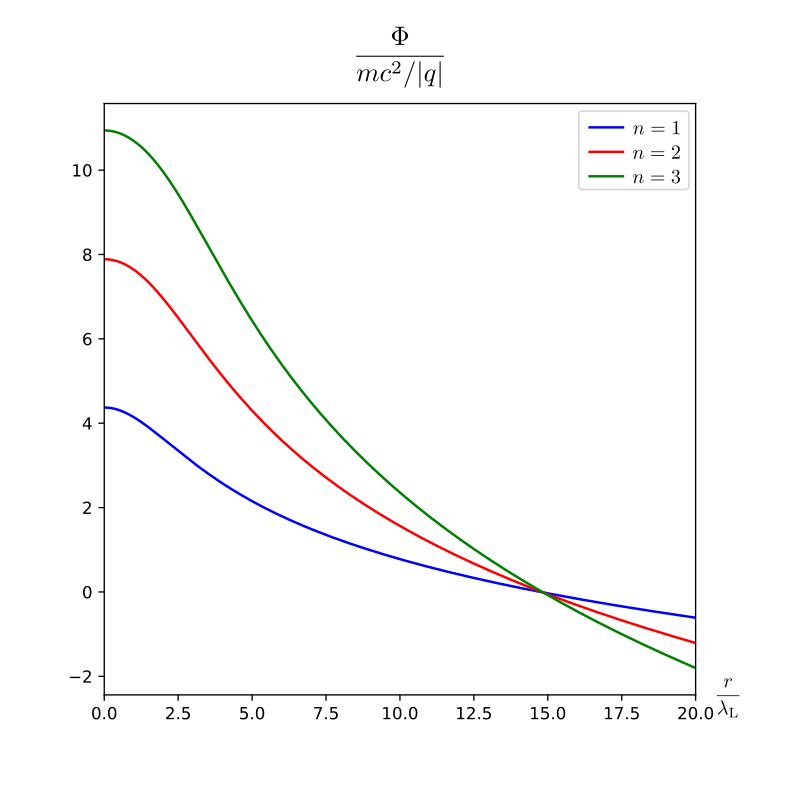}}
\vskip\baselineskip
\subfigure[]{\includegraphics[width=0.48\textwidth]{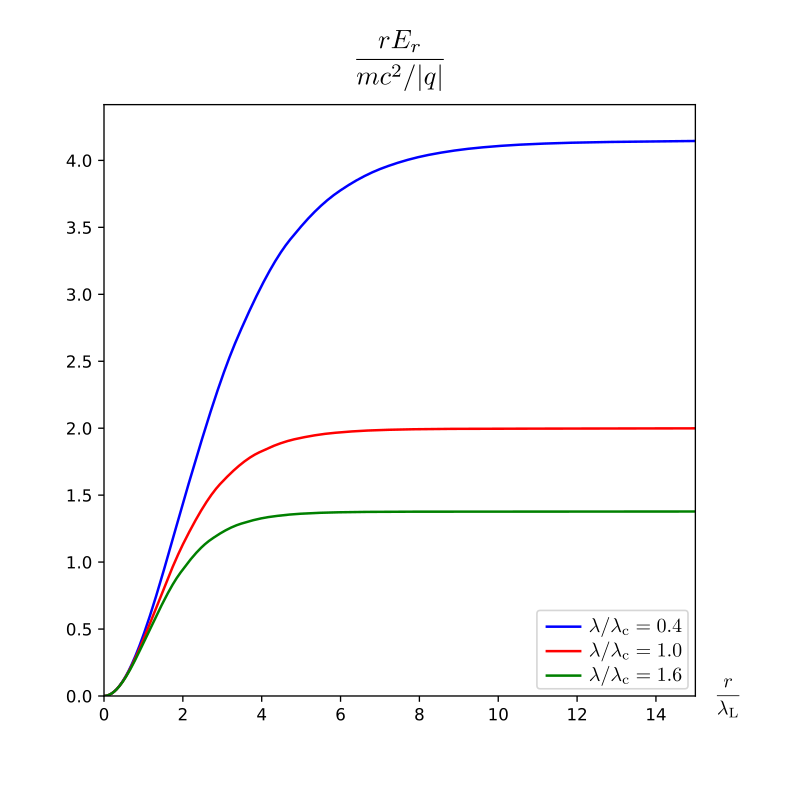}}
\hfill
\subfigure[]{\includegraphics[width=0.48\textwidth]{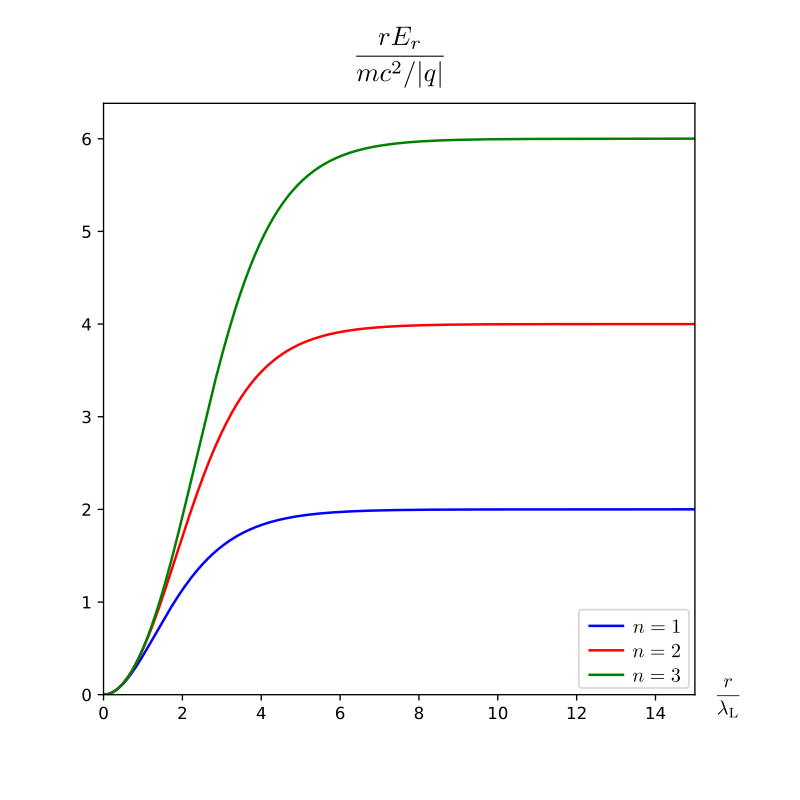}}
\caption{Profiles of the scalar potential $\Phi$ in the unit of $mc^{2}/|q|$ for (a) $ \lambda / \lambda_{\rm c} = 0.4, 1.0, 1.6 $ with $n=1$ and (b) $n=1,2,3$ with $\lambda = \lambda_{\rm c}$, and profiles of the radial component of electric field multiplied by radial distance $ r E_{r} $ in the unit of $ mc^{2}/|q| $, for (c) $ \lambda / \lambda_{\rm c} = 0.4, 1.0, 1.6 $ with $n=1$ and (d) $n=1,2,3$ with $\lambda = \lambda_{\rm c}$,
as functions of a dimensionless length variable $r / \lambda_{\rm L}$. The obtained values of electric charge per unit length along $z$ axis $ \bar{Q}_{{\rm U}(1)} / 2\pi \lambda_{\rm L}^{2} n_{\rm s} |q| $ are 1.02 for $n=1$, 2.02 for $n=2$, and 3.01 for $n=3$.}
\label{fig:405}
\end{figure}

\begin{figure}[H]
\centering
\subfigure[]{\includegraphics[width=0.48\textwidth]{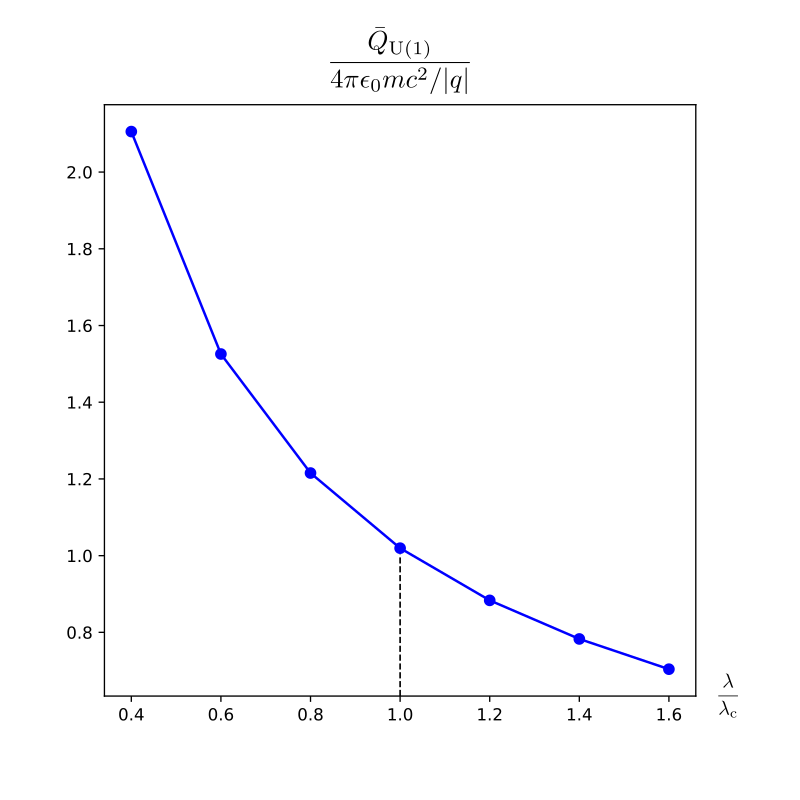}}
\hfill
\subfigure[]{\includegraphics[width=0.48\textwidth]{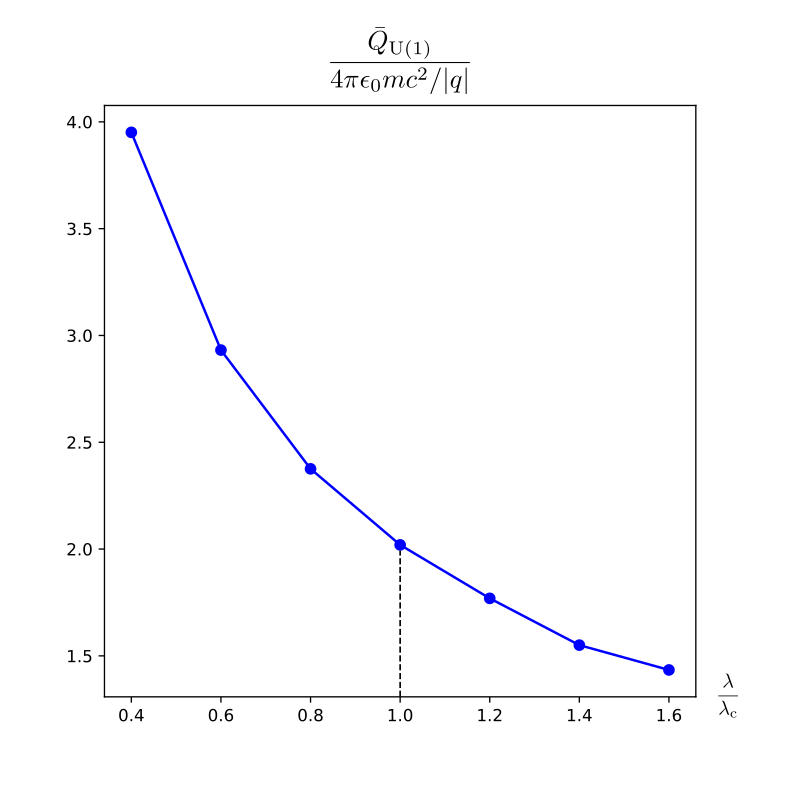}}
\caption{The electric charge of (a) $n=1$ vortices and (b) $n=2$ vortices for various values of the quartic self-interaction coupling \(\lambda\) in the unit of $ 4\pi \epsilon_{0} m c^{2} / |q| $.}
\label{fig:408}
\end{figure}

Profiles of the scalar potential $\Phi(r)$ (the neutral scalar field $N(r)$) is monotonic decreasing (increasing) and its decreasing (increasing) behavior slows down to logarithmic \eqref{432} due to accumulation of electric charge (matter) whose density \eqref{215} is localized at the vortex site as shown in figure \ref{fig:405}-(a)--(b).
This marginally confining scalar potential $\Phi$ gives the radial component of electric field \(E_{r}\) which increases from zero near the origin and decreases as the Coulomb force of \(1/r\) behavior at long range \eqref{443} irrespective of the strength of quartic self-interaction coupling $\lambda$.
Hence the electric fields of vortex solutions, multiplied by radial coordinate $r$, approach constant values at large $r$ as shown in figure \ref{fig:405}-(c)--(d).
Since the numerical value of electric charge per unit length along $z$ axis $ \bar{Q}_{\mathrm{U}(1)} $ of a vortex can be obtained by both direct spatial integration \eqref{315} and the asymptotic value from each graph of $r E_{r}$, comparison of the integrated charge values and asymptotic values of $r E_{r}$ from numerical data shows less than $1\%$ error, another validity criterion of our numerical analysis.
As quartic self-interaction coupling $\lambda$ increases, magnitude of the electric charge per unit length along $z$ axis $ \bar{Q}_{\mathrm{U}(1)} $ decreases monotonically as noticed in figure \ref{fig:405}-(c), and this decreasing behavior is confirmed in figure \ref{fig:408}.
In weak coupling regime, there is chance to discover this vortex carrying electric charge larger than its critical value
\begin{align}
\bar{Q}_{\text{U}(1)} \ge
\frac{4\pi\epsilon_{0} m c^{2}}{|q|}
= 5.69 \times 10^{-5} \text{C/m}
.
\end{align}
At the critical coupling $\lambda = \lambda_{\rm c}$ \eqref{312}, the BPS bound \eqref{421} holds and comparison of a Bogomolny equation \eqref{427} and charge density $\rho$ \eqref{215} leads to the proportionality between the magnetic flux \eqref{402} and the electric charge per unit length along $z$ axis \eqref{315}.
Equal spacings of the curves of $r E_{r}$ at large $r$ in figure \ref{fig:405}-(d) confirm this BPS property \eqref{434} numerically with error less than $1\%$.
In other words, it means that, if superconducting materials have huge quartic self-interaction coupling $\lambda$, the charged vortex of such type I$\!$I superconducting samples can become less distinguishable from the neutral Abrikosov-Nielsen-Olesen vortex.
Even in the regime of extremely strong coupling regime, it does not logically rule out the possibility of finding an electric charge carried by an $n=1$ vortex is less than the minimum unit fundamental charge of an electron per unit length along $z$ axis in an extremely thin superconducting sample.
A rough order estimation of $\big[ \bar{Q}_{\text{U}(1)} \times \text{\AA} \text{ order thickness} \big]/e \sim \mathcal{O}(10^{4})$ requires at least an additional suppression factor of the order $10^{-5}$ from the dimensionless strong coupling $\lambda/\lambda_{\rm c}$.

Once charged vortices are generated, nonzero Coulombic electric field is formed.
However, the electrostatic repulsive force exerted on charged objects, e.g., Cooper pairs and electrons, is exactly canceled by the attractive force by gapless acoustic phonon. Hence perfect conductivity is sustained in the critical coupling $g=g_{\rm c}$ despite of nonzero electric field.
If there exists a charge carrier which couples to electromagnetism and phonon interaction with different couplings, its current can probe imperfect conductivity in the superconducting material of consideration.

Since the static solitonic extended objects of our consideration have nonzero momentum density
\begin{align} 
T\indices{^t_i} = \epsilon_{0} \epsilon^{ij} E^{j} B + \frac{m}{q} j^{i} +  q n_{\rm s} A^{i}
,
\end{align}
the vortex solutions accompanied with both electric and magnetic field are usually carry the angular momentum per unit length along $z$ axis $J$,
\begingroup
\allowdisplaybreaks
\begin{align}
J 
&
= \int d^{2} \boldsymbol{x} \,\mathcal{J} 
\nonumber\\
&
= \int d^{2} \boldsymbol{x}\, 
\epsilon^{ij} \Big[ \epsilon_{0} x^{k} (\partial_{k} \Phi) \partial_{i} A_{j} - q  x_{i} (|\Psi|^{2} - n_{\rm s}) \Big( A_{j} - \frac{\hbar}{q} \partial_{j} \Omega \Big) + \hbar n_{\rm s} \partial_{j} (x_{i} \Omega) \Big]
\nonumber\\
&
= \int d^{2} \boldsymbol{x}\, 
\Big[ \epsilon_{0} x^{k} (\partial_{k} \Phi) B - q \epsilon^{ij} x_{i} (|\Psi|^{2} - n_{\rm s}) \Big( A_{j} - \frac{\hbar}{q} \partial_{j} \Omega \Big) \Big]
- \hbar n_{\rm s} n \int d^{2} \boldsymbol{x}
,
\label{406}
\end{align}
\endgroup
where angular momentum density $\mathcal{J}$ in $xy$ plane is
\begin{align}
\mathcal{J} = \epsilon^{ij} x_{i} T\indices{^t_j} = \epsilon^{ij} x_{i} \Big[\epsilon_{0} \epsilon^{jk} E_{k} B + \frac{m}{q} j^{j} + q n_{\rm s} A^{j}\Big]
.\label{416}
\end{align}
The last divergent term in \eqref{406} is computed under the assumption of complex scalar phase for static $n$ vortices located at arbitrary positions $ \boldsymbol{x}_{a} = ( x_{a}, y_{a} ) \in \mathbb{R}^{2} $ ($a = 1, 2, \cdots, n$), valid up to redundant gauge degree of freedom,
\begin{align}
\Omega = \sum_{a=1}^{n} \tan^{-1} \frac{y-y_{a}}{x-x_{a}}
. \label{433}
\end{align}
Even though the angular momentum is proportional to vorticity $n$, the obtained divergent term $(\hbar/2)(-2n_{\rm s})(\text{volume of sample}) n$ in \eqref{406} is not the total spin carried by $n$ topological vortices superimposed at the center but can naturally be interpreted as the total spin consisting of the contribution from all $2n_{{\rm s}}$ constant background electrons of unit spin $\hbar/2$  in the form of Cooper pairs over the volume of the sample felt by the vortex of vorticity $n$. It may mean that all the background electrons are perfectly aligned throughout the whole sample by an unknown mechanism of back reaction.
Therefore, physically meaningful angular momentum of a vortex per unit length along $z$ axis $J$ \eqref{406} may be the finite first part after removing the second divergent part, $J \propto \hbar n_{\rm s} n \int d^{2} \boldsymbol{x}$.

\begin{figure}[H]
\centering
\subfigure[]{\includegraphics[width=0.48\textwidth]{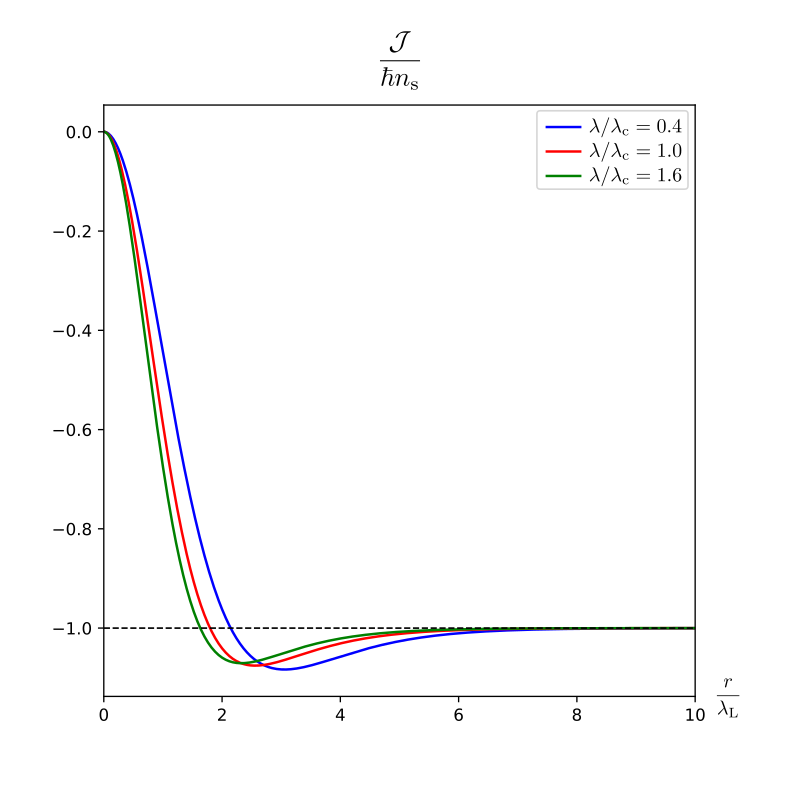}}
\hfill
\subfigure[]{\includegraphics[width=0.48\textwidth]{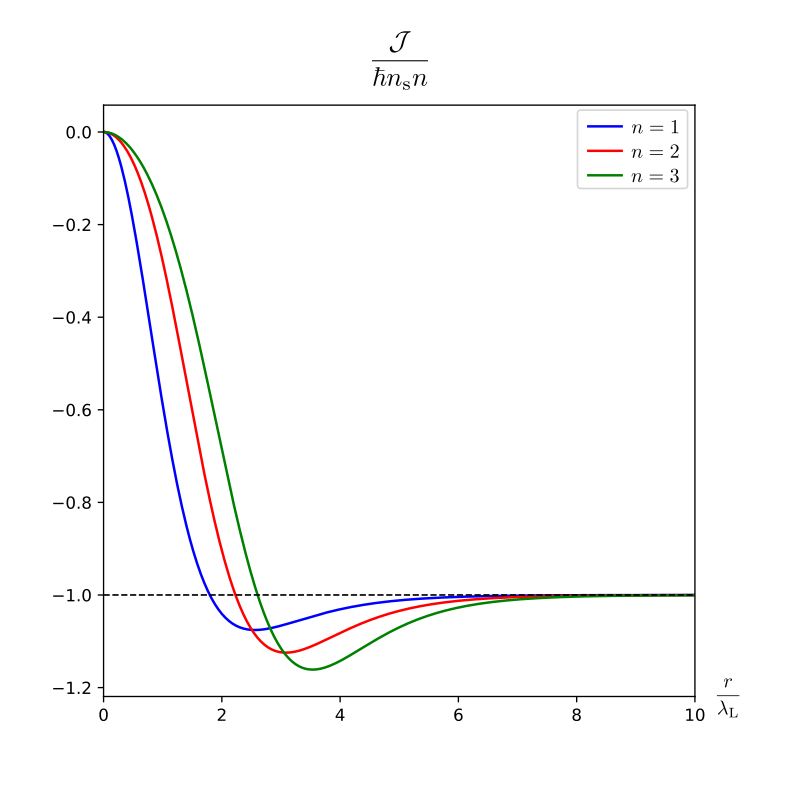}}
\caption{Profiles of the angular momentum density \( \mathcal{J} \) in the unit of $ \hbar n_{\rm s} n $, for (a) $\lambda/\lambda_{\rm c} = 0.4,~1.0,~1.6$ of $n=1$ vortex and (b) $n=1,~2,~3$ with $\lambda/\lambda_{\rm c} = 1$, as functions of a dimensionless length parameter \(r / \lambda_{\rm L}\).}
\label{fig:409}
\end{figure}

As plotted in figure \ref{fig:409}-(a)--(b), the graphs of angular momentum density \(\mathcal{J}\) \eqref{416} for cylindrically symmetric configurations, after being subtracted by the nonzero constant density $-\hbar n_{\rm s} n$ denoted by the horizontal dotted lines, commonly have a positive peak at the origin and a ring shape with negative minimum at a finite radius.
Hence the charged vortex consists of two inner and outer regions which are spinning in opposite directions.
For cylindrically symmetric vortices of vorticity $n$, finite piece of the angular momentum \eqref{416} is calculated to be zero by use of the Gauss' law \eqref{209},
\begin{align}
J + \hbar n_{\rm s} n \int d^{2} \boldsymbol{x}
&=
2\pi \hbar \int_{0}^{\infty} dr\, r 
\bigg[ \frac{\epsilon_{0}}{q} \frac{d\Phi}{dr} \frac{dA}{dr} 
- (A+n) (|\Psi|^{2} - n_{\rm s}) \bigg]
= 0.
\label{431}
\end{align}
Thus the obtained charged vortices are interpreted to be left as spinless, i.e., a contribution from coexistence of the radial component of electric field $E_r$ and the magnetic field along $z$ axis $B$ is exactly cancelled by the sum of the current density term and the constant background charge density term at least for cylindrically symmetric configurations.

In the BPS limit of critical coupling, it is noted that the finite part of angular momentum is expressed solely by either the gauge field,
\begin{align}
J \mp \frac{\bar{Q}_{{\rm U}(1)}}{2} n = \epsilon_{0} \int d^{2} \boldsymbol{x}\, x^{i} \Big( \delta^{ij} E^{j} B - \frac{qn_{\rm s}}{\epsilon_{0}} \epsilon^{ij}  A_{j} \Big) ,
\end{align}
or by the two scalar fields,
\begin{align}
J \mp \frac{\bar{Q}_{{\rm U}(1)}}{2} n = \frac{q n_{\rm s} \Phi_{\rm L}}{2} \int d^{2} \boldsymbol{x}\, x^{i} \bigg[ \frac{\sqrt{\epsilon_{0}}}{q \lambda_{{\rm L}}^{2} n_{\rm s}^{2}} (n_{\rm s} - |\Psi|^{2}) \partial_{i} N \pm \partial_{i} \ln \frac{|\Psi|^{2}}{\displaystyle \prod_{a=1}^{n} |\boldsymbol{x} - \boldsymbol{x}_{a}|^{2}} \bigg] .
\end{align}

If the static objects, specifically the straight vortex strings along $z$ axis and equivalently the point-like topological vortices in the $xy$ plane, carry both the magnetic flux $\Phi_{B}$ \eqref{402} and the electric charge $Q_{{\rm U}(1)}$ \eqref{228}, each charged topological vortex with the magnetic flux in the current field theoretic model of superconductivity forms a composite object named the fluxon \cite{Wilczek:1981du} which is applied for the theory of anyon superconductivity \cite{Chen:1989xs}.
A representative vortex example of fluxon is the Chern-Simons vortex for which the Gauss' law with time-independent fields gives nonzero scalar potential, $ A^{0} \propto |\Psi|^{2} - v^{2} \ne 0 $, in the BPS limit \cite{Hong:1990yh, Jackiw:1990aw}. However, these spinless fluxons, the charged spinless vortices of consideration, cannot be regarded as a species of anyon of arbitrary spin in $(1+2)$ dimensions \cite{Wilczek:1982wy}. Nevertheless these spinless vortices of vorticity $n$ can probe the constant background spin density $ (\hbar / 2) (-2n_{\rm s}) $.

The topological vortices classified by the quantized magnetic flux \eqref{402} have nonzero current density $j^{i} \neq 0$ \eqref{202} and hence carry additionally the nonzero magnetic dipole moment,
\begin{align}
m_{z} = \frac{1}{2} \int d^{2} x \, \epsilon^{ij} x_{i} j^{j}
.\label{435}
\end{align}
This  magnetic moment is computed exactly in the BPS limit by a quantized value for the $n$ vortices of arbitrary separations,
\begin{align}
m_{z}
= - \frac{ 2\pi \hbar \epsilon_{0} c^{2}}{q} |n|
= \frac{\hbar}{2m} \bar{Q}_{{\rm U}(1)}
.
\end{align}

A characteristic of nonrelativistic dynamics distinguished from relativistic dynamics appears in comparison of energy flux density
\begin{align}
- T\indices{^i_t} = \epsilon_{0} c^{2} (\boldsymbol{E} \times \boldsymbol{B})^{i} - \frac{\hbar^2}{2m} (\overline{\mathcal{D}_{i} \Psi} \mathcal{D}_{t} \Psi + \overline{\mathcal{D}_{t} \Psi} \mathcal{D}_{i} \Psi) - \partial_{i} N \partial_{t} N 
,\label{219}
\end{align}
and momentum density
\begin{align}
T\indices{^t_i} = 
\epsilon_{0} (\boldsymbol{E} \times \boldsymbol{B})^{i} 
+ \frac{m}{q} j^{i}
- \frac{1}{v_{N}^{2}} \partial_{t} N \partial_{i} N + q n_{\rm s} A^{i}
. \label{220}
\end{align}
For nontrivial configurations away from the homogeneous vacuum, the Galilean boost symmetry for matter part of the action \eqref{201} instead of the Lorentz boost symmetry makes the energy flux density and the momentum density different, $-T\indices{^i_t} \neq c^{2} T\indices{^t_i}$.
Stress components $T\indices{^i_j}$ of the energy-momentum tensor are symmetric $T\indices{^i_j} = T\indices{^j_i}$ with the help of the equations of motion due to the symmetry of spatial rotation,
\begingroup
\allowdisplaybreaks
\begin{align}
T\indices{^i_j} =&\, \Big[ \frac{\epsilon_0}{2} (\boldsymbol{E}^{2} + c^{2}\boldsymbol{B}^{2}) + \frac{i \hbar}{2}(\bar{\Psi} \mathcal{D}_{t} \Psi  
- \overline{\mathcal{D}_{t} \Psi} \Psi ) 
- \frac{\hbar^{2}}{2m}|\mathcal{D}_{k}\Psi|^{2} 
+ \frac{1}{2v_{N}^{2}}(\partial_{t}N)^{2} -\frac{1}{2}(\partial_{k}N)^{2} \nonumber\\
&\,~~ 
- \lambda (|\Psi|^{2} - v^{2}) \Big( |\Psi|^{2} + \frac{g}{\lambda} N - v^{2} \Big) + q \Phi n_{\rm s} \Big] \delta_{ij} \nonumber\\
&\, 
- \epsilon_{0} E^{i} E^{j} - \epsilon_{0} c^{2} B^{i} B^{j} 
+ \frac{\hbar^{2}}{2m}(\overline{\mathcal{D}_{i} \Psi} \mathcal{D}_{j} \Psi 
+ \overline{\mathcal{D}_{j} \Psi} \mathcal{D}_{i} \Psi) 
+ \partial_{i} N \partial_{j} N
.\label{444}
\end{align}
\endgroup
Though it vanishes everywhere for the BPS objects as explained in \eqref{420}, the stress components \eqref{444} usually have nonzero pressure and shear force for arbitrary static vortex configurations.

A perfect cancellation of static interactions in tree level between the long ranged Coulomb repulsion and the long ranged attraction by gapless acoustic phonon is obtained at the critical cubic Yukawa type coupling $g=g_{c}$ \eqref{327} \cite{NRAH}.
In this section, we show that nonsingular vortex solutions also exist only at the same critical coupling $ g = g_{\rm c} $.
The obtained vortices of finite energy carry both quantized magnetic flux and nonzero electric charge, but zero spin.
By the nonzero electric charge, our vortices supported by the effective field theory of action \eqref{201} are different from the neutral Abrikosov-Nielsen-Olesen vortices \cite{Abrikosov:1956sx, Nielsen:1973cs}.
Even in the presence of the Coulomb repulsion between two charged vortices, the net interaction between the two charged vortices is attractive in weak coupling regime $ 0 < \lambda < \lambda_{\rm c} $ and repulsive in strong coupling regime $ \lambda > \lambda_{\rm c} $.
At the critical coupling $\lambda = \lambda_{\rm c}$ of the BPS limit, they are noninteracting.
After perfect cancellation of the Coulomb repulsion and the attraction by gapless acoustic phonon, residual net interaction is decided by the $1$-loop level competition of an attractive Higgs interaction and a repulsive magnetic interaction.
This classification exactly coincides with the well-known type I and I$\!$I superconductors made first in \cite{Abrikosov:1956sx}.

Though our analysis relies much on numerical works and the objects are studied under cylindrical symmetry in this section, mathematically rigorous analysis is necessary for this field theory at the borderline of type I and I$\!$I superconductivity, which is expected to be beneficial for studying multi-centered vortex solutions of arbitrary shape in this BPS limit \cite{Jeon:2024jbs}.
In addition, stability of the vortices classified by the winding number $n$ called the vorticity is tested in relation with the flat direction along the vacuum configuration of neutral scalar field $N=0$.

Finally, vortex-antivortex configurations of zero net electric charge and zero magnetic flux are particularly intriguing as the global $\mathrm{U}(1)$ vortex-antivortex configurations did in the Kosterlitz-Thouless theory \cite{Kosterlitz:1973xp}, while we have considered only the configurations consisting of vortices in this paper.
These electrically neutral configurations are free from the harmful logarithmic divergence of scalar potential at long range even away from the critical coupling $g\ne g_{\rm c}$.
Since these vortex-antivortex configurations lead to nonzero electric dipole moment, the signal of such vortex-antivortex phase may easily be detected in experiments.

\section{Conclusion and Discussion}

An effective field theory of the proposed action involving a $\text{U}(1)$ gauge field of electromagnetism, a Schr\"odinger type complex scalar field of Cooper pair, and a gapless neutral scalar field of acoustic phonon is examined, in which spontaneous breakdown of $\text{U}(1)$ symmetry is initiated by constant background charge density, and static vortex solutions are investigated.
The regular vortex solutions are supported only in the critical cubic Yukawa type coupling between complex and neutral scalar fields, $g=g_{\rm c}$, different from the homogeneous Higgs vacuum of superconducting phase, allowed for arbitrary cubic Yukawa type coupling $g$.
The obtained vortices characterized by the quantized magnetic flux also carry electric charge but spinless.
Despite of the same profiles of static complex scalar field and vector potential, our charged vortices are different from the neutral Abrikosov-Nielsen-Olesen vortices of the Ginzburg-Landau theory \cite{Abrikosov:1956sx} and of the relativistic and nonrelativistic Abelian Higgs model \cite{Nielsen:1973cs, Kim:2024gfn, Jeon:2025snd}.
The critical coupling $g=g_{\rm c}$ means a critical role of the acoustic phonon to generate the vortices of finite energy in the superconducting state.
The finite energy per unit length along $z$ axis for each obtained charged vortex is achieved through nonperturbative perfect cancellation of the long ranged Coulomb repulsion between two charged vortices and the long ranged attractive force mediated by gapless neutral scalar field of acoustic phonon.

In the context of interaction between the charged vortices, energetics of multi-vortices forces that they are attractive in the weak coupling regime of quartic self-interaction coupling $0 < \lambda < \lambda_{\rm c}$ which corresponds to type I superconductivity and repulsive in its strong coupling regime $\lambda > \lambda_{\rm c}$ which corresponds to type I$\!$I superconductivity.
The borderline of type I and I$\!$I superconductors is shown to be the BPS limit of these charged vortices \eqref{421} whose arbitrary separation does not cost additional energy.
Therefore, the discovery of BPS limit for these charged vortices is the nonperturbative confirmation of static interaction balance between the short ranged repulsion mediated by massive mode of the $\text{U}(1)$ gauge field and the short ranged attraction mediated by massive Higgs degree of the complex scalar field discussed in perturbative scheme \cite{NRAH}.
This classification of type I and I$\!$I superconductors exactly coincides with the conventional categorization of $s$-wave superconductors for the Abrikosov-Nielsen-Olesen vortices \cite{Abrikosov:1956sx, Bogomolny:1975de}.

Though the obtained vortices carrying simultaneously electric charge and magnetic flux are fluxons \cite{Wilczek:1981du}, these spinless vortices are not anyons of fractional spin \cite{Wilczek:1982wy} obeying fractional statistics \cite{Arovas:1985yb, Wu:1984py}.
In relation with spin, our spinless vortices with electric charge and magnetic flux whose gauge dynamics is governed solely by the Maxwell term are distinguished from the spinning Chern-Simons vortices with both $\mathrm{U}(1)$ charge and magnetic flux \cite{Hong:1990yh, Jackiw:1990aw}, which have been studied for anyon superconductivity \cite{Chen:1989xs}.
In order to detect our charged vortex configuration, either they are compared with the observed data of the known charged vortices in experiments or new experiments will be designed with adequate superconducting samples \cite{Careri:1965, Warlimonthandbook}.
Until now, theoretical predictions \cite{Khomskii:1995, Jan:2001} and experimental observations \cite{Hagen:1991, Blatter:1996, Nagaoka:1998, Kumagai:2001, Mounce:2011, Sahu:2022} of charged vortices have been made in high temperature superconductors, in which the vortices mostly become charged by trapping charged particles.
Since the charge of a vortex obtained from the action \eqref{201} is generic without trapped charges, it is intriguing to devise the experiments which can distinguish the generic charge from the trapped charges.

Once we have the Ginzburg-Landau theory described in terms of the Ginzburg-Landau free energy whose constituents are the electromagnetic fields written by scalar and vector potentials and a complex scalar order parameter \cite{Ginzburg:1950sr, Abrikosov:1956sx}, the first question in the context of field theory is what the corresponding field theory is.
If we begin with the well-known Abelian Higgs model, every static properties, e.g., identification of the Ginzburg-Landau free energy as the energy of static configurations, the superconducting vacuum and the neutral Abrikosov-Nielsen-Olesen vortices with classification of type I and I$\!$I superconductors, are reproduced successfully without added fields.
However, the Abelian Higgs model with relativistic complex scalar field $\phi$ with the light speed as the characteristic propagation speed is hard to be a field theory of conventional superconductivity of low critical temperature, e.g., $T_{\rm c} \sim$ a few Kelvin.
If we subsequently consider the nonrelativistic Abelian Higgs model by a simple replacement from the light speed $c$ to a nonrelativistic propagation speed $v_{\rm p}$, $v_{\rm p} / c \ll 1$, all the aforementioned properties are again fulfilled \cite{Jeon:2025snd}.
On the other hand, there remains a homework to identify the existence and value of such characteristic speed $v_{\rm p}$ of the Cooper pair.
If nonrelativistic limit of the Abelian Higgs model is taken by $\phi = e^{-\frac{i}{\hbar} m c^{2}} \Psi$, the nonrelativistic action is nothing but the effective action \eqref{465} with the Schr\"odinger type complex scalar field $\Psi$ but without the neutral scalar field, $N=0$ or $g=0$.
As we discussed in the section 4.1, this field theory can not support regular static vortex solutions of finite energy.
Even if they exist, the Gauss' law forbids neutral vortex but dictates charged vortex of infinite energy.
Thus the proposed effective field theory of critical cubic Yukawa type coupling including a gapless neutral scalar field of acoustic phonon \cite{FTSC, NRAH} seems to be the only viable nonrelativistic field theory of $s$-wave superconductivity \cite{Nagaosa:1999ud, Wen:2004ym, Bennemann:2008, Coleman:2015, Arovas:2019}.
Since its vortices are not neutral but charged, it is unclear until now whether the effective action \eqref{301} describes the conventional superconductivity explained by the Ginzburg-Landau theory or not.\footnote{In fact, the Ginzburg-Landau theory does not specify the form of charge density in the expected Gauss' law because of the absence of time component of the covariant derivative term of complex scalar field in the Ginzburg-Landau free energy.}

Since our analysis on the vortices are limited to cylindrically symmetric solutions except for the BPS limit of critical quartic self-interaction coupling $ \lambda = \lambda_{\rm c} $ and the studies rely heavily on numerical method, the plausible argument on the nonexistence of nonsingular vortex solutions away from the critical cubic Yukawa type coupling $ g = g_{\rm c} $ lacks mathematical rigor.
This mathematical issues are worth tackling in order to conclude definitely on existence or nonexistence of charged vortex solutions of $ g = g_{\rm c} $ including multi-vortices with arbitrary separation and vortex lattices.

In the viewpoint of semiclassical approximation, physical quantities of the vortices are found by solving the classical equations of motion and the BPS limit holds in leading level.
Therefore, further quantum corrections are necessarily taken into account without \cite{Abrikosov:1975a} and with time dependence \cite{NRAH}.
The subleading corrections may include some $1$-loop contributions \cite{Rajaraman:1982is} and the method utilizing collective coordinates is beneficial \cite{AlonsoIzquierdo:2024nbn, Gervais:1975yg}.

In relativistic field theories, existence of the BPS bounds is understood through their supersymmetric extensions and the $\mathcal{N}=2$ supersymmetric field theory of the Abelian Higgs model with the critical coupling $\lambda = \lambda_{\rm c}$ is achieved \cite{DiVecchia:1977nxl, Witten:1978mh}.
It would be intriguing to find supersymmetric extension of the effective action \eqref{301} at the critical quartic self-interaction coupling by overcoming the difficulty due to nonrelativistic matters.

Recent development of the inhomogeneous Abelian Higgs model encourages construction of the inhomogeneous version of our field theory with the addition of electric and magnetic impurity terms \cite{Hook:2013yda, 2013iqa}, which support inhomogeneous BPS and nonBPS vacuum and vortices \cite{Kim:2024gfn, Jeon:2024jbs, Jeon:2025snd, Ashcroft:2018gkp}.
The existence of inhomogeneous vacuum seems universal as shown also in the Chern-Simons Higgs model \cite{Bazeia:2024fgo, Kim:2024gpu}.
Possible origin of magnetic impurity term in its supersymmetric extension is identified by the Fayet-Illiopoulos term \cite{Kim:2023abp}.
Effect of inhomogeneity is expected to support inhomogeneous BPS and nonBPS vacuum and affects much on the properties of vortices in the inhomogeneous version of our effective field theory containing constant electric impurity as a generic constituent \cite{NRIAH}.

Our analysis on the vortices deals only with static properties including energy, charge, magnetic flux, and spin with their conservation.
Effective field theory of the proposed action \eqref{201} contains a relativistic $\text{U}(1)$ gauge field of electromagnetism with the fastest propagation speed $c$ and two nonrelativistic complex and neutral scalar fields without and with a slow propagation speed $v_{N}$.
In the framework of perturbative quantum field theory, huge speed difference, $v_{N}/c \sim 10^{-5}$, breaks interaction balance away from static limit and provides a plausible explanation on temperature-fragile nature and low critical temperature of $s$-wave superconductors \cite{NRAH}.
Another breakdown of static interaction balance between the electromagnetic repulsion with the light speed $c$ and the nonrelativistic Higgs attraction without characteristic speed can lead to fuzziness of the borderline of type I and I$\!$I superconductors.
Therefore, huge speed difference foretells that classical and quantum dynamics of these charged vortices are expected to be nontrivial.
Recent progress on soliton dynamics is suggestive \cite{Alonso-Izquierdo:2025suz, Krusch:2024vuy} and gives different results from $90^{\circ}$ degree low scattering for the relativistic Abrikosov-Nielsen-Olesen vortices \cite{Manton:1981mp, Manton:2004tk, Weinberg:2012pjx}.
Low energy classical scattering of our two or more vortices carrying both electric charge and magnetic flux is expected to be intriguing as has been done for other charged vortices \cite{Kim:1992yz}.
Such unusual motion of vortices in superconductor may be measured by the flux-flow Hall effect near future \cite{Bardeen:1965zz, Hagen:1991, Nagaoka:1998, vanOtterlo:1995zz}.

\section*{Acknowledgement}

The authors would like appreciate Kwang-Yong Choi, Chanyong Hwang, Chanju Kim, O-Kab Kwon, Hyunwoo Lee, Tae-Ho Park, and  D. D. Tolla for discussions on various topics of condensed matter physics and field theory.
This work was supported by the National Research Foundation of Korea(NRF) grant with grant number NRF-2022R1F1A1073053 and RS-2019-NR040081 (Y.K.).

\end{document}